\documentclass[twocolumn]{autart}    

\usepackage{graphicx}          

\usepackage{natbib}
\usepackage{algorithm,algorithmic}
\usepackage{amsmath,amssymb,amsfonts}
\usepackage{graphicx}
\usepackage{textcomp}
\usepackage{mathtools,mathbbol,dsfont}
\usepackage{bm}
\usepackage{textcomp}
\usepackage[all]{xy}
\usepackage{chemarrow}
\usepackage{color}
\usepackage{mathrsfs,array}
\usepackage{booktabs}
\usepackage[version=4]{mhchem}
\usepackage{enumerate}
\usepackage{tikz}
\usepackage{wrapfig}
\usepackage{multirow}

\newtheorem{theorem}{Theorem}[section]
\newtheorem{corollary}{Corollary}[section]
\newtheorem{lemma}{Lemma}[section]
\newtheorem{definition}{Definition}[section]

\newtheorem{remark}{Remark}[section]
\newtheorem{example}{Example}[section]

\begin{document}

\begin{frontmatter}

\title{Structure-conditioned input-to-state stability for layer-by-layer molecular computations in parallel chemical reaction networks\thanksref{footnoteinfo}} 

\thanks[footnoteinfo]{The earlier version of this paper will be presented at the 23rd IFAC World Congress. This work was funded by the National Nature Science Foundation of China under Grant No. 12320101001, and the National Foreign Expert Project of China under Grant No. S20250211. (Corresponding author: Chuanhou Gao.)}

\author[First]{Renlei Jiang}\ead{jiangrl@zju.edu.cn},    
\author[First,First2]{Chuanhou Gao}\ead{gaochou@zju.edu.cn},               
\author[Second]{Denis Dochain}\ead{denis.dochain@uclouvain.be}  

\address[First]{School of Mathematical Sciences, Zhejiang University, Hangzhou 310058, China.}  
\address[First2]{Center for interdisciplinary Applied Mathematics, Zhejiang University, Hangzhou, 310058, China.}             
\address[Second]{ICTEAM, UCLouvain, B\^{a}timent Euler, avenue Georges Lema\^{i}tre 4-6, 1348 Louvain-la-Neuve, Belgium}        


\begin{abstract}                          
Molecular computation in chemical reaction networks (CRNs) now constitutes a foundational framework for designing programmable biological systems. However, prevailing design methodologies primarily treat parallelism of chemical reactions as a liability, consequently motivating researchers to redirect research focus toward leveraging parallelism to implement layer-by-layer computations of composite functions in coupled mass-action systems (MASs). MASs exhibiting this property are termed composable. Present composability verification for MASs mainly depends on input-to-state stability (ISS) conditions, with structural characteristics of networks remaining underexplored. This paper investigates the structural conditions under which two MASs are composable. By leveraging ISS-Lyapunov functions, we identify a class of CRN architectures, whose reduced systems have zero deficiency, that guarantee composability with other networks. We also extend our conclusions to encompass some CRN architectures possessing nonzero deficiency. Some examples are presented to demonstrate the validity of our theoretical results. Finally, we employ our methods to devise an algorithm for constructing MASs capable of executing specified molecular computations.
\end{abstract}

\end{frontmatter}

\section{Introduction}
In recent years, research focused on the programmability of biochemical systems through molecular computations has garnered substantial attention, representing an integration of multiple disciplinary fields including computer science, mathematics, and systems biology. This emerging field uses molecular components, such as DNA and enzymes, as computational elements to design systems capable of processing information, making decisions, and executing complex behaviors within biological environments, which is known as ``molecular computer''. Unlike conventional silicon-based computers, molecular computers offer advantages including low energy consumption, parallel computing, and biocompatibility, making them particularly suitable for applications such as medical diagnosis \citep{zhang2020cancer} and data storage \citep{wang2023parallel}.

Chemical reaction network (CRN) serves as a primary framework for molecular computations. They characterize interactions between chemical species through reactions and are widely used to represent complex biochemical processes such as metabolism and gene regulation. From a mathematical perspective, CRN governed by mass-action kinetics, termed mass-action system (MAS), can be modeled by a system of ordinary differential equations (ODEs) that describe the temporal evolution of species concentrations. The structure (depending on reactions) and parameters (reaction rate constants) of a MAS determine its dynamical behavior, including properties such as stability \citep{feinberg1987chemical}, persistence \citep{craciun2013persistence}, and oscillations \citep{domijan2009bistability}. This versatility establishes CRNs as powerful tools for understanding both natural and engineered chemical systems, while also providing a robust foundation for molecular computations.

The conceptual foundation of molecular computation dates back to the early 1960s, when \cite{jacob1961genetic} proposed that biomolecules could execute conditional statements, just like programming languages, to control protein production in bacteria. About three decades later, \cite{adleman1994molecular} employed DNA molecules to solve the Hamiltonian Path Problem (a well-known NP-hard problem), which was the first real example of molecular computation. Utilizing the inherent parallelism of chemical reactions, DNA molecules in test tubes can yield computational results within a few minutes. Subsequently, a growing number of experiments have implemented more sophisticated molecular computing systems capable of addressing more complex computational tasks \citep{liu2000dna,cherry2018scaling,chen2024synthetic}. Furthermore, \cite{soloveichik2010dna} demonstrated that any CRN endowed with mass-action kinetics can be physically implemented via DNA strand displacement, establishing that abstract CRN models are always realizable as tangible DNA molecules. This discovery has attracted substantial research interest in theoretical investigations of molecular computation, prompting dedicated efforts toward developing chemical reaction systems with learning and reasoning capabilities. \cite{vasic2020crn++} introduced a novel language termed CRN++ for programming deterministic CRNs to execute computations, thereby contributing to the establishment of a comprehensive molecular programming framework. \cite{chen2023rate} employed a specialized class of CRNs, known as rate-independent CRNs, to implement computations of piecewise linear continuous functions. \cite{anderson2025chemical} conceptualized CRNs as analog computers and systematically investigated the computational speed at which these CRNs perform calculations. In addition, several investigations have explored methodologies for implementing neural networks using CRNs \citep{vasic2022programming,fan2025automatic_1}.

Although significant advances have been achieved in CRN-based molecular computation, parallelism is often regarded as a drawback rather than an advantage in these studies due to the inherent conflict between the parallel nature of chemical reactions and the sequential execution requirements of conventional computing paradigms. Current research predominantly employed chemical oscillators to control the occurrence sequence of reactions \citep{shi2025controlling}, thereby adapting molecular systems to the serialized execution mechanism required by conventional computational tasks (for example, the implementation of a fully-connected nonlinear neural network in \citep{fan2025automatic_1}). However, this approach requires adding a number of additional oscillatory species (32 oscillatory species in \citep{fan2025automatic_1}), while simultaneously introducing computational errors \citep{fan2025automatic_2}. Consequently, it is imperative to investigate strategies that harness the inherent parallelism of chemical reactions for molecular computations, enabling the concurrent execution of multiple reactions to achieve desired computational results. \cite{chalk2019composable} formalized this concept as composability, which means that the computation of a composite function can be achieved by interconnecting CRNs that compute elementary functions. They investigated the composability of deterministic rate-independent CRNs and precisely characterized the class of functions computable by such CRNs. \cite{severson2019composable} extended this line of research to stochastic rate-independent CRNs. \cite{jiang2025input} extended the concept of composability to MASs and developed a composability verification method based on input-to-state stability (ISS). In contrast to the severely constrained computational capacity of rate-independent CRNs \citep{chen2023rate}, MASs have been proven to be Turing-universal \citep{fages2017strong}, implying that any computation can be embedded into a class of polynomial ODEs and implemented by CRNs. Consequently, research on the composability of MAS facilitates the realization of a broader spectrum of computational functions. However, few studies have investigated the relationship between ISS and CRN structural properties \citep{chaves2005input}, making it difficult to infer composability directly from network architecture.

Based on the definitions and conclusions established in \citep{jiang2025input} and our previous work on connection between network structure and composability \citep{jiang2025structure}, this paper continues to investigate the specific CRN structure that enables the layer-by-layer computation of composite function through network composition. The main contributions are listed below:
\begin{itemize}
    \item We establish connections between ISS-Lyapunov functions, persistence, and global asymptotic stability (\textit{Theorem} \ref{thm_main} and \textit{Corollary} \ref{coro_global_stable}, also presented in \citep{jiang2025structure});
    \item We establish composability for a class of structurally specific mass-action chemical reaction computers (msCRCs) whose reduced systems are weakly reversible and have zero deficiency (\textit{Theorem} \ref{thm_deficiency_zero} and \textit{Corollary} \ref{coro_deficiency_zero}, relax the conditions for the same results in \citep{jiang2025structure});
    \item We extend this composability to some networks with non-zero deficiency, such as birth-death processes (\textit{Theorem} \ref{thm_bd}) and CRNs with $\dim \mathscr{S}=1$ (\textit{Theorem} \ref{thm_dim1});
    \item We leverage above theoretical results to design an algorithm that constructs a msCRC to perform molecular computation of composite functions involving polynomial root-solving (\textit{Algorithm} \ref{algorithm}).
\end{itemize}

Our paper is organized as follows. Some necessary preliminaries on CRNs are given in Section \ref{section_2}. Section \ref{section_3} is used to revisit the definitions and conclusions from \citep{jiang2025input}, and establish connections between ISS-Lyapunov functions, persistence, and global asymptotic stability. In Section \ref{section_4}, we investigate the composability of msCRCs whose reduced systems are weakly reversible and have zero deficiency. We also extend this composability to some networks with nonzero deficiency in Section \ref{section_5}. Finally, Section \ref{section_6} is dedicated to the conclusion of the whole paper.

\noindent{\textbf{Mathematical Notation:}}\\
\rule[1ex]{\columnwidth}{0.8pt}
\begin{description}
\item[\hspace{0em}{$\mathbb{R}^n, \mathbb{R}^n_{\geq 0},\mathbb{R}^n_{>0},\mathbb{Z}^n_{\ge 0}$}]: $n$-dimensional real space; $n$-dimensional non-negative real space; $n$-dimensional positive real space; $n$-dimensional non-negative integer space.
\item[\hspace{0em}{$s^{v_{\cdot j}}$}]: $s^{v_{\cdot j}}\triangleq\prod_{i=1}^{n}s_{i}^{v_{ij}}$, where $s\in\mathbb{R}^{n},~v_{\cdot j} \in \mathbb{Z}^n$, and $0^0=1$.
\item[\hspace{0em}{$x/y$}]: $x/y\triangleq\left(\frac{x_1}{y_1},\cdots ,\frac{x_n}{y_n}\right)^{\top}$, where $x,y\in\mathbb{R}^{n}_{>0}$.
\item[\hspace{0em}{$\mathrm{Ln}(x)$}]: $\mathrm{Ln}(x)\triangleq\left(\ln(x_1),\cdots , \ln(x_n)\right)^{\top}$, where $x\in\mathbb{R}^{n}_{>0}$.
\item[\hspace{0em}{$\mathcal{K},\mathcal{K}_{\infty},\mathcal{KL}$}]: $\gamma : \mathbb{R}_{\ge 0} \to \mathbb{R}_{\ge 0}$ is a class $\mathcal{K}$ function if it is continuous strictly increasing and satisfies $\gamma(0)=0$; $\gamma$ is a class $\mathcal{K}_\infty$ function if it is a class $\mathcal{K}$ function and satisfies $\lim_{s \to \infty}\gamma(s)=\infty$; $\beta : \mathbb{R}_{\ge 0} \times \mathbb{R}_{\ge 0} \to \mathbb{R}_{\ge 0}$ is a class $\mathcal{KL}$ function if for each fixed $t$ the mapping $\beta(\cdot,t)$ is a class $\mathcal{K}$ function and for each fixed $s$ it decreases to zero on $t$ as $t \to \infty$.
\end{description}
\rule[1ex]{\columnwidth}{0.8pt}


\section{Preliminaries}\label{section_2}
In this section, we present some basic concepts related to CRNs \citep{feinberg1987chemical}.

Consider a CRN with $n$ species, denoted by $S_1,...,S_n$, and $r$ reactions with the $j$th reaction written as
$$\sum_{i=1}^nv_{ij}S_i\to \sum_{i=1}^nv'_{ij}S_i,$$
where $v_{.j}, v'_{.j}\in\mathbb{Z}_{\geq 0}^n$ represent the reactant complex and the product complex of the reaction, respectively. For simplicity, this reaction is often written as $v_{.j}\to v'_{.j}$. We thus present the related notions in the CRN theory.  

\begin{definition}[CRN]
    A CRN consists of three finite sets: 
    \begin{enumerate}
        \item a finite \textit{species} set $\mathcal{S}=\{ S_1,...,S_n\}$;
        \item a finite \textit{complex} set $\mathcal{C}=\bigcup_{j=1}^r{\left\{ v_{\cdot j},v_{\cdot j}^{\prime} \right\}}$, where the $i$th entry of $v_{\cdot j}$, i.e., $v_{ij}$, represents the stoichiometric coefficient of species $S_i$ in complex $v_{\cdot j}$;
        \item a finite \textit{reaction} set $\mathcal{R}=\bigcup_{j=1}^r{\left\{ v_{\cdot j}\rightarrow v_{\cdot j}^{\prime} \right\}}$ satisfying
        \begin{enumerate}
            \item $\forall v_{\cdot j}\rightarrow v_{\cdot j}^{\prime}\in \mathcal{R}, v_{\cdot j}\ne v_{\cdot j}^{\prime}$,
            \item $\forall v_{\cdot j} \in \mathcal{C}, \exists v_{\cdot j}^{\prime} \in \mathcal{C}$ such that $v_{\cdot j}\rightarrow v_{\cdot j}^{\prime}\in \mathcal{R}$ or $v_{\cdot j}^{\prime}\rightarrow v_{\cdot j}\in \mathcal{R}$.
        \end{enumerate}
    \end{enumerate}
 The triple $(\mathcal{S},\mathcal{C},\mathcal{R})$ is usually used to express a CRN.
\end{definition}

\begin{definition}[Stoichiometric subspace]\label{def_stoichiometric_subspace}
    For a $(\mathcal{S},\mathcal{C},\mathcal{R})$, the linear subspace $\mathscr{S}=\textrm{span}\{ v_{\cdot 1}^{\prime}-v_{\cdot 1},...,v_{\cdot r}^{\prime}-v_{\cdot r}\}$ is called the \textit{stoichiometric subspace} of the network. The dimension of $\mathscr{S}$, denoted by $\dim \mathscr{S}$ is called the dimension of the CRN $(\mathcal{S},\mathcal{C},\mathcal{R})$.
\end{definition}

\begin{definition}[Stoichiometric compatibility class]\label{def_stoichiometric compatibility class}
For a $(\mathcal{S},\mathcal{C},\mathcal{R})$, let $s_0 \in \mathbb{R}_{\ge0}^n$, the set $s_0+\mathscr{S}=\{s_0+s|s \in \mathscr{S}\}$ is called the \textit{stoichiometric compatibility class} of $s_0$. Further, $(s_0+\mathscr{S}) \bigcap \mathbb{R}_{\ge 0}^n$ is called the \textit{nonnegative stoichiometric compatibility class} of $s_0$, and $(s_0+\mathscr{S}) \bigcap \mathbb{R}_{> 0}^n$ is called the \textit{positive stoichiometric compatibility class} of $s_0$.
\end{definition}

\begin{example}\label{example1}
    A CRN follows
    \begin{align}\label{additionCRN}
        S_{1} &\longrightarrow S_{1} + S_{3}\ , &   
        S_{2} &\longrightarrow S_{2} + S_{3}\ ,  &           S_{3} &\longrightarrow \varnothing,  
    \end{align}
    where the last reaction refers to an outflow reaction. The species set is $\mathcal{S}=\{S_1,S_2,S_3\}$, the complex set is $\mathcal{C}=\{(1,0,0)^\top,(1,0,1)^\top,(0,1,0)^\top,(0,1,1)^\top,(0,0,1)^\top,(0,0, \\ 0)^\top \}$, and the stoichiometric subspace is $\mathscr{S}=\textrm{span}\{(0,0,\\1)^\top,(0,0,-1)^\top\}$.
\end{example}

When a CRN is equipped with mass-action kinetics, the rate of reaction $v_{\cdot j} \to v_{\cdot j}^{\prime}$ is measured by $k_js^{v_{\cdot j}}$, where $k_j>0$ represents the rate constant, and $s \in \mathbb{R}_{\ge 0}^n$ with each element $s_i~(i=1,...,n)$ to represent the concentration of the species $S_i$.

\begin{definition}[MAS]\label{def_MAS}
    Let $\kappa=(k_1,\cdots ,k_r)$ be the set of reaction rate constants, where $k_j$ represents the rate constant for reaction $v_{\cdot j} \to v_{\cdot j}^{\prime},~1\le j \le r$. A \textit{MAS} is a CRN $(\mathcal{S},\mathcal{C},\mathcal{R})$ taken with a mass-action kinetics $\kappa$, often labeled by $(\mathcal{S},\mathcal{C},\mathcal{R},\kappa)$. A \textit{generalized MAS} is labeled by $(\mathcal{S},\mathcal{C},\mathcal{R},\kappa(t))$ where $\kappa$ is not always constant but time-varying.
\end{definition}

The dynamics of $(\mathcal{S},\mathcal{C},\mathcal{R},\kappa)$ describes the change of concentrations of all species over time $t$, and thus follows
\begin{equation}\label{general dynamics}
    \frac{\mathrm{d}s(t)}{\mathrm{d}t}=\sum_{j=1}^{r}k_js^{v_{\cdot j}}\left(v_{\cdot j}^{\prime}-v_{\cdot j} \right),
\end{equation}
which is essentially polynomial ODEs. By integrating (\ref{general dynamics}) from $0$ to $t$, we get
\begin{equation}\label{integ_dynamics}
 s(t) = s_0 + \sum_{j=1}^r \left(v_{\cdot j}^{\prime}-v_{\cdot j} \right) \int_0^t k_js^{v_{\cdot j}}(\tau)\mathrm{d}\tau,   
\end{equation}
where $s_0=s(0)$ is the initial state of $(\mathcal{S},\mathcal{C},\mathcal{R},\kappa)$. It is clear that the state of $(\mathcal{S},\mathcal{C},\mathcal{R},\kappa)$ will evolve in the nonnegative stoichiometric compatibility class of $s_0$, i.e., in $(s_0+\mathscr{S}) \bigcap \mathbb{R}_{\ge 0}^n$. 

We also take the CRN in (\ref{additionCRN}) as an example. By setting reaction rate $\kappa=(1,1,1)$ we have the dynamics
\begin{equation}\label{ex1_dynamics}
\frac{\mathrm{d} s_{1}}{\mathrm{d} t} =\frac{\mathrm{d} s_{2}}{\mathrm{d} t} = 0,\quad \frac{\mathrm{d} s_{3}}{\mathrm{d} t} = s_{1} + s_{2} - s_{3}.
\end{equation}

\begin{definition}[Balanced MAS]
     For a MAS $(\mathcal{S},\mathcal{C},\mathcal{R},\kappa)$ governed by (\ref{general dynamics}), a constant vector $\bar{s} \in \mathbb{R}^n_{>0}$ is called a \textit{positive equilibrium} of the system if
    \begin{equation}
        \sum_{j=1}^{r}k_j\bar{s}^{v_{\cdot j}}\left(v_{\cdot j}^{\prime}-v_{\cdot j} \right)=0.
    \end{equation}
    A MAS that admits a positive equilibrium is said to be a \textit{balanced MAS}.
\end{definition}

\begin{definition}[Complex balanced MAS]\label{cbMAS}
    For a MAS $(\mathcal{S},\mathcal{C},\mathcal{R},\kappa)$ governed by (\ref{general dynamics}), a positive equilibrium $\bar{s}$ is called a \textit{complex balanced equilibrium} of the system if
    \begin{equation}
        \sum_{\{j|v_{\cdot j}=z\}}k_j\bar{s}^{v_{\cdot j}}=\sum_{\{j|v_{\cdot j }^{\prime}=z\}}k_j\bar{s}^{v_{\cdot j}},\quad \forall z \in \mathcal{C}.
    \end{equation}
    A MAS that admits a complex balanced equilibrium is said to be a \textit{complex balanced MAS}. If a MAS has a complex balanced equilibrium, then all of the other equilibria (if any) are complex balanced. 
\end{definition}

\begin{definition}[Stability \citep{rouche1977stability}]\label{stability}
    The equilibrium point $\bar{s}$ of system (\ref{general dynamics}) is
    \begin{enumerate}
        \item stable if for any $\varepsilon >0$, there is $\delta=\delta(\varepsilon)$ such that
        $$\left \|s(0)-\bar{s} \right \|<\delta  \Rightarrow \left \|s(t)-\bar{s} \right\|<\varepsilon, \quad \forall t \ge 0;$$
        \item asymptotically stable if it is stable and $\delta$ can be chosen such that
        $$\left \|s(0)-\bar{s} \right \|<\delta  \Rightarrow \lim_{t \to \infty}s(t)=\bar{s};$$
        \item globally asymptotically stable if 
        $$\lim_{t \to \infty}s(t)=\bar{s}, \quad \forall s(0) \in \mathbb{R}_{\ge 0}^n.$$
    \end{enumerate}
\end{definition}

\begin{definition}[Persistence \citep{craciun2013persistence}]
    Let $s(t) \in \mathbb{R}_{\ge 0}^n$ be the solution to the system (\ref{general dynamics}) with initial condition $s(0)=s_0$. We say the trajectory $s(t)$ is \textit{persistent} if
    \begin{equation*}
        \liminf_{t \to \infty} s_j(t) >0, \quad j=1,2,\cdots ,n.
    \end{equation*}
\end{definition}

The structure of CRNs has been a focal point in research, as CRNs with specific architectures can exhibit particular dynamical behaviors in their corresponding MAS (regardless of rate constants values $\kappa$).

\begin{definition}[Weakly reversible CRN]
    A CRN $(\mathcal{S},\mathcal{C},\mathcal{R})$ is 
    \begin{enumerate}
        \item \textit{reversible} if $\forall v_{\cdot j} \to v_{\cdot j}^{\prime} \in \mathcal{R}$ it holds $v_{\cdot j}^{\prime} \to v_{\cdot j} \in \mathcal{R}$;
        \item \textit{weakly reversible} if $\forall v_{\cdot j} \to v_{\cdot j}^{\prime} \in \mathcal{R}$, there exists a sequence of complexes $v_{\cdot j_1},\cdots ,v_{\cdot j_p} \in \mathcal{C}$ such that $v_{\cdot j}^{\prime} \to v_{\cdot j_1} \in \mathcal{R}, v_{\cdot j_1} \to v_{\cdot j_2} \in \mathcal{R}$, $\cdots$, $v_{\cdot j_{p-1}} \to v_{\cdot j_p} \in \mathcal{R}$, $v_{\cdot j_p} \to v_{\cdot j} \in \mathcal{R}$. 
    \end{enumerate}
\end{definition}

 A CRN can be seen as a directed graph, where the nodes represent complexes and the directed edges correspond to reactions. Each connected component of the directed graph is called a \textit{linkage class}. From the perspective of directed graph, a CRN is weakly reversible if and only if each of its linkage classes is strongly connected, i.e., every edge of this linkage class is a part of an oriented cycle. Also, combining this notion with Definition \ref{cbMAS} suggests that a complex balanced MAS must be weakly reversible.

\begin{definition}[Deficiency]
    For a CRN $(\mathcal{S},\mathcal{C},\mathcal{R})$, let $c$ denote the number of complexes and $\ell$ denote the number of linkage classes. The \textit{deficiency} of the CRN is defined by $\delta =c-\ell-\dim \mathscr{S}$, where $\mathscr{S}$ is defined in \textit{Definition} \ref{def_stoichiometric_subspace}.
\end{definition}

\begin{example}\label{ex_deficiency}
    Consider the following CRN
    \begin{equation}\label{CRN_deficiency}
        \begin{tikzpicture}
	    \node (a) at (0,0) {$Z_1$};
	    \node (b) at (2,0) {$Z_2$};
	    \node (c) at (1,1.7) {$Z_3$};
	    \draw[->] (a) -- (b) node[midway, below] {};
    	\draw[->] (b) -- (c) node[midway, right] {};
    	\draw[->] (c) -- (a) node[midway, left] {};
        \end{tikzpicture}
    \end{equation}
    which is weakly reversible but not reversible. In addition, it has three complexes $(1,0,0)^\top$, $(0,1,0)^\top$ and $(0,0,1)^{\top}$, one linkage class, and its stoichiometric subspace is $\mathscr{S}=\{(-1,1,0)^{\top},(0,-1,1)^{\top},(1,0,-1)^{\top}\}$. Hence, its deficiency is $\delta=3-1-2=0$.
\end{example}

Note that the deficiency of a CRN is nonnegative because it can be seen as the dimension of a certain linear subspace \citep{feinberg2019foundations}. A CRN with low deficiency may have hundreds of species and hundreds of reactions, but its kinetic behavior can be directly determined. For example, see the following lemma or Deficiency One Theorem (both of them can be found in \citep{feinberg1987chemical}).

\begin{lemma}[Deficiency Zero Theorem]\label{lem_deficiency_zero}
    Suppose a MAS $(\mathcal{S},\mathcal{C},\mathcal{R},\kappa)$ is weakly reversible and has zero deficiency. Then for any $\kappa \in \mathbb{R}^{r}_{>0}$, the MAS is complex balanced. Moreover, there exists within each positive stoichiometric compatibility class precisely one equilibrium $\bar{s}=\bar{s}(s_0,\kappa)$, and the pseudo-Helmholtz free energy function
    \begin{equation}\label{Lyafun_deficiency_zero}
        V(s,\bar{s})=\sum_{j=1}^r \left(s_j(\ln s_j-\ln \bar{s}_j-1)-\bar{s}_j\right),~s \in \mathbb{R}^{n}_{> 0}
    \end{equation}
    is a Lyapunov function that ensures $\bar{s}$ to be locally asymptotically stable.
\end{lemma}

\begin{definition}[Strongly endotactic CRN]
    A CRN $(\mathcal{S},\mathcal{C},\mathcal{R})$ is
    \begin{enumerate}
        \item $w$-endotactic with respect to a certain $w \in \mathbb{R}^n$ if for the set of all reactants such that the reaction vectors are not orthogonal to $w$, denoted by $\mathbb{V}_w$, it holds
        $$w^{\top}\left(v_{\cdot j}^{\prime}-v_{\cdot j}\right)<0$$
        for $v_{\cdot j} \in \mathbb{V}_w$ that satisfies $w^{\top}\left(v_{\cdot j}-v\right)\ge 0,~\forall v \in \mathbb{V}_w$, and
        $$w^{\top}\left(v_{\cdot k}^{\prime}-v_{\cdot k}\right)>0$$
        for $v_{\cdot k} \in \mathbb{V}_w$ that satisfies $w^{\top}\left(v_{\cdot k}-v\right)\le 0,~\forall v \in \mathbb{V}_w$;
        \item endotactic if it is $w$-endotactic with respect to any $w \in \mathbb{R}^n$;
        \item strongly endotactic if it is endotactic and for every vector $w \in \mathbb{R}^n$ not orthogonal to the stoichiometric subspace of $(\mathcal{S},\mathcal{C},\mathcal{R})$, there exists a reaction $v_{\cdot j} \to v_{\cdot j}^{\prime}$ such that
        \begin{enumerate}
            \item $w^{\top}\left(v_{\cdot j}^{\prime}-v_{\cdot j}\right)<0$;
            \item $w^{\top}\left(v_{\cdot j}-v_{\cdot k}\right)\ge 0,~k=1,2,\cdots,r$.
        \end{enumerate}
    \end{enumerate}
\end{definition}

The above definition is a structural characterization of CRNs using algebraic language by \cite{gopalkrishnan2014geometric}. Further, they established that any strongly endotactic CRN is persistent.


\section{ISS-based msCRCs Composition}\label{section_3}
In this section, we revisit concepts from \citep{jiang2025input} related to ISS-based composable msCRCs, and further establish connections between dynamic composability conditions and ISS-Lyapunov functions.

\subsection{Composable MASs for molecular computation}
In the study of molecular computation, MAS serves as a prevalent mathematical framework designed to represent computational inputs via species concentrations and to encode computational outputs as the limiting steady state of the relevant species concentrations.

\begin{definition}[msCRC]
    A msCRC is a tuple $\mathscr{C}=(\mathcal{S},\mathcal{C},\mathcal{R},\kappa,\mathcal{X},\mathcal{Y})$, where $(\mathcal{S,C,R},\kappa)$ is a MAS, $\mathcal{X} \subset \mathcal{S}$ is the input species set, and $\mathcal{Y}=\mathcal{S} \setminus \mathcal{X}$ is the output species set.
\end{definition}

\begin{definition}[Dynamic computation]\label{Def_dynamic_compute} Given a msCRC $\mathscr{C}=(\mathcal{S},\mathcal{C},\mathcal{R},\kappa,\mathcal{X},\mathcal{Y})$ with $m~(m<n)$ species in $\mathcal{X}$, and a positive function $\sigma:\mathbb{R}^{m}_{\ge 0} \to \mathbb{R}^{n-m}_{\ge 0}$, denote the dynamics of this msCRC by
    \begin{equation}\label{CRC_dynamics}
    \begin{cases}
        \dot{x}=f(x,y), \\
        \dot{y}=g(x,y),
    \end{cases} x(0)=x_0,y(0)=y_0,
    \end{equation}
where $x \in \mathbb{R}^{m}_{\ge 0}$ and $y \in \mathbb{R}^{n-m}_{\ge 0}$ are the concentration vectors of input species and output species, respectively. The msCRC is a dynamic computation of function $\sigma$, if the output species concentrations satisfy
    \begin{equation}\label{eq:dyCom}
  \lim_{t \to \infty}y(t)=\sigma(x_0).      
    \end{equation}
\end{definition}

\begin{remark}
    The dynamics of (\ref{CRC_dynamics}) is an alternative of the original form of (\ref{general dynamics}), which suggests that $f:\mathbb{R}_{\ge 0}^m \times \mathbb{R}_{\ge 0}^{n-m} \to \mathbb{R}_{\ge 0}^m,~g:\mathbb{R}_{\ge 0}^m \times \mathbb{R}_{\ge 0}^{n-m} \to \mathbb{R}_{\ge 0}^{n-m}$ are with the structure of polynomial functions. 
\end{remark}

Revisit \textit{Example} \ref{example1} and the corresponding dynamics (\ref{ex1_dynamics}), we can get $s_1(t)=s_1(0), s_2(t)=s_2(0)$, and 
$s_3(t)=s_1(0)+s_2(0)+\left (s_3(0)-s_1(0)-s_2(0) \right ) e^{-t}$. Thus we have $\lim_{t \to \infty}s_3(t)=s_1(0)+s_2(0)$. This means the msCRC induced by (\ref{additionCRN}) with $\mathcal{X}=\{S_1,S_2\}$ and $\mathcal{Y}=\{S_3\}$ is a dynamic computation of the function $\sigma((s_1, s_2)^\top)=s_1+s_2$. Essentially, the network given in (\ref{additionCRN}) can perform the addition computation between any two nonnegative real numbers. 

In the above example, the limiting steady state of the output is independent of its initial point $y_0$. Next we give another example whose limiting steady state of the output depends on the specific initial output concentration.

\begin{example}\label{ex_exp}
    A msCRC follows
    \begin{align}\label{exp_CRC}
      \quad X + Y &\overset{1}{\longrightarrow} X+2Y, &  
       X &\overset{1}{\longrightarrow} \varnothing  
    \end{align}
    with input species set $\mathcal{X}=\{X\}$ and output species set $\mathcal{Y}=\{Y\}$. It is easy to get
\begin{equation*}\label{function_ex1}
\begin{cases}
        \dot{x}=-x, \\
        \dot{y}=xy;
    \end{cases} \Rightarrow \quad \begin{cases}
        x(t)=x_0e^{-t}, \\
        y(t)=y_0e^{x_0\left(1-e^{-t}\right)}.
    \end{cases}
 \end{equation*}
We thus have $\lim_{t \to \infty}y(t)=y_0e^{x_0}$. This indicates that the msCRC is a dynamic computation of $\sigma(x)=e^{x}$ if the initial concentration $y_0=1$.
\end{example}

For molecular computation based on MAS, a critical issue involves investigating the dynamic composability of multiple simple molecular computing systems. Specifically, it is essential to explore the possibility of leveraging their combinations to achieve the corresponding composite computations.

\begin{definition}[msCRCs composability]\label{def_composition}
    Given two msCRCs $\mathscr{C}^\mathtt{1}=(\mathcal{S}^\mathtt{1},\mathcal{C}^\mathtt{1},\mathcal{R}^\mathtt{1},\kappa^\mathtt{1},\mathcal{X}^\mathtt{1},\mathcal{Y}^\mathtt{1})$ and $\mathscr{C}^\mathtt{2}=(\mathcal{S}^{\mathtt{2}},\mathcal{C}^{\mathtt{2}},\mathcal{R}^{\mathtt{2}},\kappa^{\mathtt{2}},\mathcal{X}^{\mathtt{2}}, \mathcal{Y}^{\mathtt{2}})$ satisfying $\mathcal{Y}^\mathtt{1} =\mathcal{X}^{\mathtt{2}}$ and $\mathcal{Y}^{\mathtt{2}}\cap \mathcal{X}^{\mathtt{1}}=\varnothing$, $\mathscr{C}^\mathtt{1}$ and $\mathscr{C}^{\mathtt{2}}$ are said to be composable, and their composition is the msCRC $\mathscr{C}^{\mathtt{2}\circ \mathtt{1}}=(\mathcal{S}^\mathtt{1} \cup \mathcal{S}^{\mathtt{2}}, \mathcal{C}^\mathtt{1} \cup \mathcal{C}^{\mathtt{2}}, \mathcal{R}^\mathtt{1} \cup \mathcal{R}^{\mathtt{2}}, \kappa^\mathtt{1} \cup \kappa^{\mathtt{2}},\mathcal{X}^\mathtt{1} \cup \mathcal{X}^{\mathtt{2}}, \mathcal{Y}^{\mathtt{2}})$.
\end{definition}

\begin{definition}[msCRCs dynamic composability]\label{Def_dynamic_composable}
Let $\mathscr{C}^\mathtt{1}=(\mathcal{S}^\mathtt{1},\mathcal{C}^\mathtt{1},\mathcal{R}^\mathtt{1},\kappa^\mathtt{1},\mathcal{X}^\mathtt{1},\mathcal{Y}^\mathtt{1})$ and $\mathscr{C}^\mathtt{2}=(\mathcal{S}^{\mathtt{2}},\mathcal{C}^{\mathtt{2}},\mathcal{R}^{\mathtt{2}},\kappa^{\mathtt{2}}, \\ \mathcal{X}^{\mathtt{2}},\mathcal{Y}^{\mathtt{2}})$ be composable, and they satisfy the following assumptions:
\begin{enumerate}
   \setlength{\itemindent}{2em}
    \item[(A.1)] $\mathscr{C}^\mathtt{1}$ and $\mathscr{C}^\mathtt{2}$ are with respective dynamics to be
    \begin{equation}\label{dx=f}
        \begin{cases}
          \dot{x}^\mathtt{1}=f^\mathtt{1}(x^\mathtt{1},y^\mathtt{1}), \\
          \dot{y}^\mathtt{1}=g^\mathtt{1}(x^\mathtt{1},y^\mathtt{1}),
        \end{cases}x^\mathtt{1}(0)=x^\mathtt{1}_0,~y^\mathtt{1}(0)=y^\mathtt{1}_0
    \end{equation} 
    and
    \begin{equation}\label{dy=g}
        \begin{cases}
            \dot{x}^\mathtt{2}=0, \\
            \dot{y}^\mathtt{2}=g^\mathtt{2}(x^\mathtt{2},y^\mathtt{2}),
        \end{cases} x^\mathtt{2}(0)=\bar{y}^\mathtt{1},~y^\mathtt{2}(0)=y^\mathtt{2}_0;
    \end{equation}
    \item[(A.2)] $\lim_{t \to \infty} y^\mathtt{1}(t)= \bar{y}^\mathtt{1},~\lim_{t \to \infty} y^{\mathtt{2}}(t)= \bar{y}^{\mathtt{2}}$.
\end{enumerate}
If the solution of the coupled system 
\begin{equation}\label{coupled_system}
    \begin{cases}
     \dot{x}^\mathtt{1}=f^\mathtt{1}(x^\mathtt{1},y^\mathtt{1}), \\
     \dot{y}^\mathtt{1}=g^\mathtt{1}(x^\mathtt{1},y^\mathtt{1}),\\
     \dot{y}^\mathtt{2}=g^\mathtt{2}(y^\mathtt{1},y^\mathtt{2}), 
    \end{cases} \begin{array}{l}
         x^\mathtt{1}(0)=x^\mathtt{1}_0,~y^\mathtt{1}(0)=y^\mathtt{1}_0, \\
         y^\mathtt{2}(0)=y^\mathtt{2}_0,
    \end{array}
\end{equation}
satisfies 
\begin{equation}\label{dynamic_composable}
    \lim_{t \rightarrow \infty} \left [ \begin{array}{c}
         y^\mathtt{1}(t)  \\
         y^\mathtt{2}(t)  \\
    \end{array} \right] =\left[ \begin{array}{c}
         \bar{y}^\mathtt{1} \\
         \bar{y}^\mathtt{2} \\
    \end{array} \right],
\end{equation}
then the two msCRCs are said to be dynamically composable.
\end{definition}
\begin{remark}
    The msCRCs composability and assumptions (A.1), (A.2) are crucial for studying their dynamic composability, which will be repeatedly referenced in the subsequent theoretical developments. If $\mathscr{C}^{\mathtt{1}}$ and $\mathscr{C}^{\mathtt{2}}$ are regarded as two msCRCs implementing distinct molecular computations, then composability implies that the output of $\mathscr{C}^{\mathtt{1}}$ serves as the input of $\mathscr{C}^{\mathtt{2}}$, and assumption (A.1) means the species in $\mathscr{C}^{\mathtt{1}}$ remain unchanged in $\mathscr{C}^{\mathtt{2}}$ (i.e., they act as \textit{catalysts}). Assumption (A.2) further indicates that the parallel $\mathscr{C}^{\mathtt{1}}$ and $\mathscr{C}^{\mathtt{2}}$ can achieve layer-by-layer computation of a composite function. The steady state of $y_1$ and $y_2$ in the coupled system (\ref{coupled_system}) can be thought as a layer-by-layer composition of their individual output limiting steady state. Therefore, if $\mathscr{C}^{\mathtt{1}}$ and $\mathscr{C}^{\mathtt{2}}$ implement the computations of functions $y^{\mathtt{1}}=\sigma_1(x^{\mathtt{1}})$ and $y^{\mathtt{2}}=\sigma_2(y^{\mathtt{1}})$, respectively, and they are dynamically composable, then their composition can serve to compute the composite function $y^{\mathtt{2}}=\sigma_2 \circ \sigma_1(x^{\mathtt{1}})$ \citep{jiang2025input}.
\end{remark}

In the following, we introduce the notion of ISS, which will play an important role in suggesting msCRCs dynamic composability.

\begin{definition}[ISS \citep{sontag1989smooth}]
 For a system 
    \begin{equation}\label{ds=f(u,s)}
        \dot{s}=f(u,s),~s(0)=s_0
    \end{equation}
    where $s(t) \in \mathbb{R}^{n}_{\ge 0},~u(t) \in \mathbb{R}^{m}_{\ge 0},~f:\mathbb{R}^{m}_{\ge 0}\times \mathbb{R}^{n}_{\ge 0} \to \mathbb{R}^n_{\ge 0}$, it is ISS regarding $(\bar{u},\bar{s})\in \mathbb{R}^m_{\ge 0} \times \mathbb{R}^n_{\ge 0}$, if there exist a class $\mathcal{KL}$ function $\beta$ and a class $\mathcal{K}$ function $\gamma$ such that for any $s_0$ and any bounded input $u(t)$, the solution $s(t)$ exists for all $t \ge 0$ and satisfiesf
    \begin{equation}\label{ISS_def}
           \left | s \left( t \right) -\bar{s} \right | \le \beta \left( \left| s_0 - \bar{s} \right| , t \right) + \gamma ( \sup \limits_{0\le \tau \le t}\left| u \left( \tau \right) -\bar{u} \right| ).
        \end{equation}   
\end{definition}

\begin{theorem}[\cite{jiang2025input}]\label{thm_ISS}
    Suppose $\mathscr{C}^\mathtt{1}$ and $\mathscr{C}^\mathtt{2}$ are composable and satisfy assumptions (A.1), (A.2). Consider the $y^\mathtt{2}$-related part of the system (\ref{dy=g}), if it is ISS regarding $(\bar{y}^{\mathtt{1}},\bar{y}^{\mathtt{2}})$, then $\mathscr{C}^\mathtt{1}$ and $\mathscr{C}^\mathtt{2}$ are dynamically composable.
\end{theorem}

The above theorem shows that ISS is a sufficient condition for dynamic composability. However, it is not easy to find functions $\beta$ and $\gamma$ satisfying (\ref{ISS_def}). In the following section, we will investigate sufficient conditions based on the structure of CRN that admit more straightforward verifications.

\subsection{ISS-Lyapunov Function}

The application of \textit{Theorem} \ref{thm_ISS} requires ISS condition. A standard method for proving ISS is to construct an appropriate ISS-Lyapunov function \citep{chaves2002state}.
\begin{definition}[ISS-Lyapunov function]\label{def_ISS-lyafun}
    Consider a generalized MAS $(\mathcal{S},\mathcal{C},\mathcal{R},u(t))$ written as
    \begin{equation}\label{generalized_MAS}
        \dot{s}=f(u,s)=\sum_{j=1}^{r}u_j(t)s^{v_{\cdot j}}\left(v_{\cdot j}^{\prime}-v_{\cdot j}\right),~ s(0)=s_0,
    \end{equation}
    where $u(t):[0,+\infty) \to \mathbb{U}$ is input function with $\mathbb{U} \subset \mathbb{R}^r_{>0}$. An \textit{ISS-Lyapunov function} with respect to the point $(\bar{u},\bar{s}) \in \mathbb{R}^r_{>0} \times \mathbb{R}^n_{> 0}$ is a continuous function $V:\mathbb{R}^{n}_{\ge 0} \to \mathbb{R}_{\ge 0}$ that satisfies:
    \begin{enumerate}[(i)]
        \item the restriction of $V$ to $\mathbb{R}^n_{>0}$ is continuously differentiable;
        \item $V(s) \to +\infty$ as $|s| \to +\infty$;
        \item $V(\bar{s})=0$ and $V(s) > 0,~s \in \mathbb{R}^n_{\ge 0} \backslash \{\bar{s}\}$;
        \item for each compact set $F \subset \mathbb{R}^n_{\ge 0}$, there exist class $\mathcal{K}_{\infty}$ functions $\alpha,\rho$ such that $$dV/dt=\nabla V^{\top}(s)f(u,s) \le -\alpha(|s-\bar{s}|)+\rho (|u-\bar{u}|)$$
        for all $u \in \mathbb{U}$ and $s \in F \cap (s_0+\mathscr{S})\cap \mathbb{R}^n_{>0}$.
    \end{enumerate}
\end{definition}
\begin{remark}\label{rem_ISS-lyafun}
    Note that the function $V(s)$ in the above definition is only required to be differentiable in the positive orthant. It is slightly different from the traditional ISS-Lyapunov function in \citep{sontag1995characterizations}, where $V$ is differentiable in $\mathbb{R}^n$. The relaxed requirements enable the application of logarithmic Lyapunov functions, which are widely used in CRN theory.
\end{remark}

\begin{theorem}\label{thm_main}
Consider a generalized MAS $(\mathcal{S},\mathcal{C},\mathcal{R},u(t))$ governed by (\ref{generalized_MAS}). Suppose $\lim_{t \to \infty} u(t) = \bar{u}$, $s(t)$ is a bounded solution to (\ref{generalized_MAS}), and $\bar{s}\in \mathbb{R}^n_{> 0}$ is the unique equilibrium of MAS $\dot{s}=f(\bar{u},s)$ in $ (s_0+\mathscr{S})\cap \mathbb{R}^n_{> 0}$. If $V$ is an ISS-Lyapunov function with respect to $(\bar{u},\bar{s})$ for (\ref{generalized_MAS}), then either $s(t) \to \bar{s}$ or $s(t) \to \partial \mathbb{R}^n_{\ge 0}$.
\end{theorem}

The detailed proof appears in Appendix A, which may apply to all lemmas, theorems and corollaries.

In fact, the traditional ISS-Lyapunov function in \citep{sontag1995characterizations} implies the ISS property, which consequently ensures $s \to \bar{s}$ (see \citep{jiang2025input}). The reason why \textit{Theorem} \ref{thm_main} fails to guarantee $s \to \bar{s}$ lies in the difference noted in \textit{Remark} \ref{rem_ISS-lyafun}. Therefore, to ensure $s \to \bar{s}$ under this condition, the exclusion of another case $s(t) \to \partial \mathbb{R}^n_{\ge 0}$ is required.

\begin{corollary}\label{coro_global_stable}
    Under the condition of \textit{Theorem} \ref{thm_main}, if $s(t)$ is persistent, then $s(t) \to \bar{s}$.
\end{corollary}

\textit{Corollary} \ref{coro_global_stable} shows that $\bar{s}$ is a globally asymptotically stable equilibrium in (\ref{generalized_MAS}), which will play a vital role in the subsequent dynamic composability analysis.


\section{Dynamic Composability of Zero-Deficiency msCRCs}\label{section_4}
In this section, we investigate the dynamic composability of zero-deficiency msCRCs, establishing the relationship between the network topological structure and dynamic composability. Before giving our main results, an additional concept requires exposition.

\begin{definition}[Reduced system]
 For a msCRC $\mathscr{C}=(\mathcal{S},\mathcal{C},\mathcal{R}, \kappa,\mathcal{X},\mathcal{Y})$, a generalized MAS $\tilde{\mathscr{C}}=(\tilde{\mathcal{S}},\tilde{\mathcal{C}},\tilde{\mathcal{R}},\tilde{\kappa}(t))$ is called a reduced system of $\mathscr{C}$ if
    \begin{enumerate}[(1)]
        \setlength{\itemindent}{2em}
        \item $\tilde{\mathcal{S}}=\mathcal{Y}$;
        \item $\tilde{\mathcal{C}}=\pi_{\mathcal{Y}}(\mathcal{C})$, where $\pi_{\mathcal{Y}}$ denotes the projection of each complex $v_{\cdot j} \in \mathcal{C}$ onto species in $\mathcal{Y}$;
        \item $\tilde{\mathcal{R}}=\{\pi_{\mathcal{Y}}(v_{\cdot j})\overset{\tilde{\kappa}_j(t)}{\longrightarrow} \pi_{\mathcal{Y}}(v_{\cdot j}^{\prime}),~j=1,...,r\}$, where the reaction rate is given by $\tilde{\kappa}_j(t)=\kappa_j \Pi_{S_i \notin \tilde{S}}s_i^{v_{ij}}(t)$.
    \end{enumerate}
\end{definition}

\begin{example}\label{ex_reduced}
    Consider a msCRC
    \begin{equation}
        X \overset{k_1}{\longrightarrow}2X, \quad X+Y \overset{k_2}{\longrightarrow}2Y, \quad Y \overset{k_3}{\longrightarrow}\varnothing
    \end{equation}
    with $\mathcal{X}=\{X\},~\mathcal{Y}=\{Y\}$, which has dynamics
    \begin{equation}\label{eq_before_reduced}
    \begin{cases}
        \dot{x}=k_1x-k_2xy, \\
        \dot{y}=k_2xy-k_3y.
    \end{cases}
    \end{equation}
    Its reduced system is
    \begin{equation}\label{CRN_reduced_system}
        \varnothing \overset{k_3}{\longleftarrow} Y \overset{k_2x(t)}{\longrightarrow}2Y
    \end{equation}
    with dynamics to be $\dot{y}=k_2x(t)y-k_3y$.
\end{example}

The reduced system is constructed by incorporating the dynamics of the input species into the rate constants, thereby removing the input species to achieve system simplification. This concept is widely used in the theoretical analysis of CRN \citep{anderson2011proof,zhang2025persistence}. It is noted that the reduced system preserves the same dynamics as the output species of the original system (for example, (\ref{CRN_reduced_system}) has the same dynamics as the $y$-related part in (\ref{eq_before_reduced})). This property enables us to analyze the dynamical behavior of the original system when coupled with other systems through the study of its reduced system. In addition, given the similarity between condition (iv) in \textit{Definition} \ref{def_ISS-lyafun} and the definition of standard Lyapunov function, the pseudo-Helmholtz free energy function (\ref{Lyafun_deficiency_zero}) may serve as a candidate ISS-Lyapunov function. Based on these we give our main results of this section, which enables direct determination of dynamic composability from the structural properties of the msCRC.

\begin{theorem}\label{thm_deficiency_zero}
    Suppose $\mathscr{C}^\mathtt{1}$ and $\mathscr{C}^\mathtt{2}$ are composable and satisfy assumptions (A.1), (A.2), and the reduced system of $\mathscr{C}^{\mathtt{2}}$, denoted by $\tilde{\mathscr{C}}^{\mathtt{2}}$, is weakly reversible and has zero deficiency. If the trajectory $y^{\mathtt{2}}(t)$ of $\tilde{\mathscr{C}}^{\mathtt{2}}$ is persistent, then $\mathscr{C}^\mathtt{1}$ and $\mathscr{C}^\mathtt{2}$ are dynamically composable.
\end{theorem}

\begin{remark}
    The proof of \textit{Theorem} \ref{thm_deficiency_zero} proceeds in two stages. First, we demonstrate that (\ref{Lyafun_deficiency_zero}) indeed constitutes an ISS-Lyapunov function for the generalized MAS $\tilde{\mathscr{C}}^{\mathtt{2}}$. Second, we employ this function to establish dynamic composability. A simplified version of the first-stage conclusion has been proven in \citep{chaves2005input}, i.e., in the case of the single linkage class, while the second-stage result remains previously unaddressed.
\end{remark}

\begin{corollary}\label{coro_deficiency_zero}
    Suppose $\mathscr{C}^\mathtt{1}$ and $\mathscr{C}^\mathtt{2}$ are composable and satisfy assumptions (A.1), (A.2), and the reduced system of $\mathscr{C}^{\mathtt{2}}$, denoted by $\tilde{\mathscr{C}}^{\mathtt{2}}$, is weakly reversible and has zero deficiency. Then $\mathscr{C}^\mathtt{1}$ and $\mathscr{C}^\mathtt{2}$ are dynamically composable if one of the following conditions is true
    \begin{enumerate}[(1)]
        \setlength{\itemindent}{1em}
        \item $\tilde{\mathscr{C}}^{\mathtt{2}}$ has single linkage class;
        \item $\tilde{\mathscr{C}}^{\mathtt{2}}$ has only two species;
        \item $\tilde{\mathscr{C}}^{\mathtt{2}}$ is with $\dim \mathscr{S}=2$;
        \item $\tilde{\mathscr{C}}^{\mathtt{2}}$ is strongly endotactic.
    \end{enumerate}
\end{corollary}

\begin{remark}
    \textit{Corollary} \ref{coro_deficiency_zero} is a generalization of \textit{Theorem 10} in \citep{jiang2025structure}, which required weak reversibility, zero deficiency, single linkage class, and mass conservation. The generalized result holds under less restrictive structural conditions.
\end{remark}

All the conditions in \textit{Corollary} \ref{coro_deficiency_zero} depend on the structure of msCRCs. Therefore, we can efficiently determine the dynamic composability by leveraging the structural properties of the msCRCs, without recourse to dynamical equations.

\begin{example}\label{ex_softmax}
    Consider a msCRC
    \begin{equation}\label{exp3CRN}
       X_i + Y_i \overset{1}{\longrightarrow} X_i+2Y_i,\quad  X_i \overset{1}{\longrightarrow} \varnothing, \quad i=1,2,3 
    \end{equation}
    with $\mathcal{X}^\mathtt{1}=\{X_1,X_2,X_3\},~\mathcal{Y}^\mathtt{1}=\{Y_1,Y_2,Y_3\}$. By \textit{Example} \ref{ex_exp} we know that this msCRC can dynamically compute $\sigma_1 (x_1,x_2,x_3)=(e^{x_1},e^{x_2},e^{x_3})^{\top}$. Consider another msCRC 
    \begin{equation}\label{normCRN}
    \begin{split}
        Z_1+Y_2+Y_3 \overset{1}{\longrightarrow} Z_2+Y_2+Y_3, \\
        Z_2+Y_3+Y_1 \overset{1}{\longrightarrow} Z_3+Y_3+Y_1, \\
        Z_3+Y_1+Y_2 \overset{1}{\longrightarrow} Z_1+Y_1+Y_2,
    \end{split}
    \end{equation}
    with $\mathcal{X}^\mathtt{2}=\{Y_1,Y_2,Y_3\},~\mathcal{Y}^\mathtt{2}=\{Z_1,Z_2,Z_3\}$, whose reduced system is 
    \begin{equation}\label{reduced_CRN1}
        \begin{tikzpicture}
	    \node (a) at (0,0) {$Z_1$};
	    \node (b) at (2,0) {$Z_2$};
	    \node (c) at (1,1.7) {$Z_3$};
	    \draw[->] (a) -- (b) node[midway, below] {$y_2(t)y_3(t)$};
    	\draw[->] (b) -- (c) node[midway, right] {$y_3(t)y_1(t)$};
    	\draw[->] (c) -- (a) node[midway, left] {$y_1(t)y_2(t)$};
        \end{tikzpicture}
    \end{equation}
    As demonstrated in \textit{Example} \ref{ex_deficiency}, (\ref{reduced_CRN1}) is weakly reversible and has zero deficiency. By \textit{Lemma} \ref{lem_deficiency_zero}, for any $\left(y_1(t),y_2(t),y_3(t)\right)\equiv \left(\bar{y}_1,\bar{y}_2,\bar{y}_3\right) \in \mathbb{R}^3_{>0}$, there exists within each positive stoichiometric compatibility class precisely one equilibrium. Specially, $$(\bar{z}_1,\bar{z}_2,\bar{z}_3)=\left(\frac{\bar{y}_1}{\bar{y}_1+\bar{y}_2+\bar{y}_3},\frac{\bar{y}_2}{\bar{y}_1+\bar{y}_2+\bar{y}_3},\frac{\bar{y}_3}{\bar{y}_1+\bar{y}_2+\bar{y}_3}\right)$$
    is the unique complex balanced equilibrium in the positive stoichiometric compatibility class $\{(y_1,y_2,y_3) \in \mathbb{R}^3_{>0}:y_1+y_2+y_3=1\}$. That is, we have
    $\lim_{t\to \infty}\left(z_1,z_2,z_3\right)=\left(\bar{z}_1,\bar{z}_2,\bar{z}_3 \right),$
    which shows that (\ref{normCRN}) is a dynamic computation of the \textit{Normalization Function}
    $$\sigma_2(y_1,y_2,y_3)=\left(\frac{y_1}{\sum_{i=1}^3y_i},\frac{y_2}{\sum_{i=1}^3y_i},\frac{y_3}{\sum_{i=1}^3y_i}\right)^{\top}.$$
    In addition, (\ref{reduced_CRN1}) has single linkage class. Thus, by \textit{Corollary} \ref{coro_deficiency_zero}, (\ref{exp3CRN}) and (\ref{normCRN}) are dynamically composable.
\end{example}

\begin{figure}[ht]
    \centering
    \includegraphics[width=0.47\textwidth]{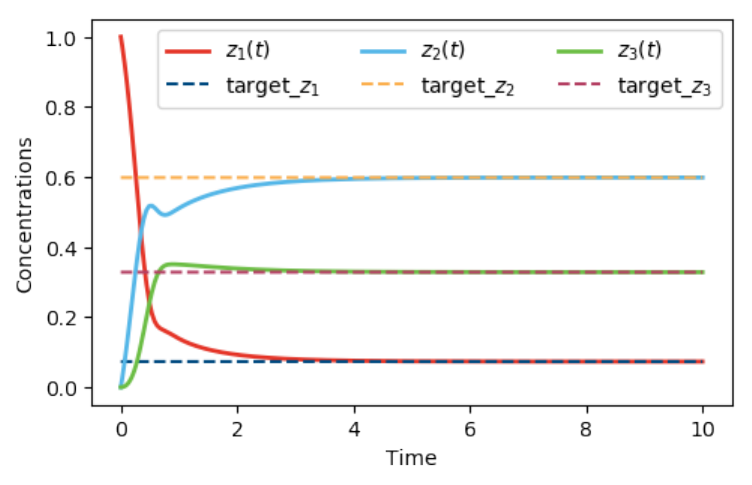}
    \caption{The numerical simulation results for (\ref{coupled_dynamics_1}) with initial values $x_1(0)=0.9,x_2(0)=3.0,x_3(0)=2.4$, $y_1(0)=y_2(0)=y_3(0)=1.0,~z_1(0)=1.0,z_2(0)=z_3(0)=0$.}
    \label{fig1}
\end{figure}

By \textit{Example} \ref{ex_softmax} we know that the msCRCs (\ref{exp3CRN}) and (\ref{normCRN}) successfully implement the computations of exponential and normalization functions, respectively. Since (\ref{exp3CRN}) and (\ref{normCRN}) are dynamically composable, their composition can serve to compute the composite function of above two functions. In other words, the composition of (\ref{exp3CRN}) and (\ref{normCRN}), with dynamics
\begin{equation}\label{coupled_dynamics_1}
    \begin{cases}
        \dot{x}_i=-x_i, \\
        \dot{y}_i=x_iy_i, \\
        \dot{z}_1=y_1y_2z_3-y_2y_3z_1, \\
        \dot{z}_2=y_2y_3z_1-y_3y_1z_2,\\
        \dot{z}_3=y_3y_1z_2-y_1y_2z_3,
    \end{cases} \quad i=1,2,3,
\end{equation}
can dynamically compute the \textit{Softmax function}
\begin{equation*}
    \sigma (x_1,x_2,x_3)=\left(\frac{e^{x_1}}{\sum_{i=1}^3e^{x_i}},\frac{e^{x_2}}{\sum_{i=1}^3e^{x_i}},\frac{e^{x_3}}{\sum_{i=1}^3e^{x_i}} \right)^{\top}.
\end{equation*}
It should be noted that the computation of the exponential function requires controlling the initial concentration $y_i(0)=1,~i=1,2,3$, while computing the normalization function necessitates satisfying $z_1(0)+z_2(0)+z_3(0)=1$ to construct the corresponding positive stoichiometric compatibility class. Consequently, when implementing the computation of composite functions through dynamics (\ref{coupled_dynamics_1}), these conditions must also be strictly satisfied. Fig. \ref{fig1} shows the numerical simulation results for (\ref{coupled_dynamics_1}), displaying exclusively the temporal evolution of concentrations of species $z_1,z_2,z_3$. As reaction time progresses, the concentrations of species $z_1,z_2,z_3$ (solid lines in the figure) converge to the corresponding computational results (dashed lines in the figure), thereby validating our conclusion. We solve the dynamical equations (\ref{coupled_dynamics_1}) using the odesolver from the scipy.integrate module in Python 3.7.0, which also apply to other simulations in this paper.

In previous molecular computation studies \citep{arredondo2022supervised,shi2025controlling}, researchers mainly viewed chemical parallelism as a drawback, introducing oscillatory signals as catalysts to control reaction sequences for compatibility with the serialized mechanism of computing composite functions . For example, to compute the Softmax function in \textit{Example} \ref{ex_softmax}, we need to design the following CRN
\begin{equation}\label{CRN_catalysts}
    \begin{split}
    O_1+O_2 &\overset{k_o}{\longrightarrow} 2O_2,\quad O_2+O_3 \overset{k_o}{\longrightarrow}2O_3,\\
    O_3+O_4 &\overset{k_o}{\longrightarrow} 2O_4, \quad O_4+O_1 \overset{k_o}{\longrightarrow} 2O_1, \\
        O_1+X_i + Y_i&\overset{1}{\longrightarrow} O_1+X_i+2Y_i, \\
        O_1+X_i &\overset{1}{\longrightarrow} O_1, \quad \quad i=1,2,3, \\
    O_3+Z_1+Y_2+Y_3 &\overset{1}{\longrightarrow} O_3+Z_2+Y_2+Y_3, \\
    O_3+Z_2+Y_3+Y_1 &\overset{1}{\longrightarrow} O_3+Z_3+Y_3+Y_1, \\
    O_3+Z_3+Y_1+Y_2 &\overset{1}{\longrightarrow} O_3+Z_1+Y_1+Y_2,
    \end{split}
\end{equation}
where $O_1,O_2,O_3,O_4$ are species generating oscillatory signals, and only $O_1,O_3$ can be used to control the occurrence sequence of reaction modules. Our methods can reduce the requirement by 4 species ($O_1,O_2,O_3,O_4$) and 4 reactions (the first 4 reactions in (\ref{CRN_catalysts})). In general, computing a $N$-layer composite function can reduce the requirement by $2N$ species and $2N$ reactions.


\section{Dynamic Composability of Nonzero-Deficiency msCRCs}\label{section_5}
In this section, we further investigate the dynamic composability of some msCRCs with non-zero deficiency.

\subsection{Dynamic composability for birth-death process}
We firstly try to establish dynamic composability for birth-death process.

\begin{definition}[Birth-death process]
  A birth-death process is a CRN $(\mathcal{S},\mathcal{C},\mathcal{R})$ satisfying
    \begin{enumerate}
        \item $\mathcal{S}=\{S\}$, i.e., the CRN is with one species only;
        \item each reaction is of the form $nS \longrightarrow n^{\prime}S$ with either $n^{\prime}=n+1$ or $n^{\prime}=n-1$. If the stoichiometric coefficient of $S$ is zero, then it corresponds to $\varnothing$.
    \end{enumerate}
\end{definition}

For convenience, under mass-action kinetics we denote the rate constants of reactions $nS \longrightarrow (n+1)S$ and $nS \longrightarrow (n-1)S$ by $\kappa_n$ and $\kappa_{-n}$, respectively. Further, we use $n_u$ to represent the largest $n$ for which $\kappa_n$ is a nonzero reaction rate constant, and similarly use $n_d$ to represent the largest $n$ for which $\kappa_{-n}$ is a nonzero rate constant.

\begin{example}
    Consider a birth-death process endowed with mass-action kinetics
    \begin{equation}
        \varnothing \ce{<=>[$\kappa_0$][$\kappa_{-1}$]}S, \quad 2S \ce{<=>[$\kappa_2$][$\kappa_{-3}$]} 3S,
    \end{equation}
   then we have $n_u=2,~n_d=3$.
\end{example}

The following two lemmas provide Lyapunov functions that can be used to prove global asymptotic stability of the equilibrium point for birth-death processes.

\begin{lemma}\label{lem_bd1}
    Consider a birth-death process endowed with mass-action kinetics $\kappa \in \mathbb{R}^r_{>0}$, described by dynamics
    \begin{equation}\label{eq_bd}
        \dot{s}=f(s)=\sum_{n \ge 0}\kappa_ns^n-\sum_{n<0}\kappa_{-n}s^{-n}.
    \end{equation}
    Suppose $\kappa_0>0$ and either of the following holds
    \begin{enumerate}
        \item $n_d>n_u$;
        \item $n_d=n_u$ and $\kappa_{-n_d}>\kappa_{n_u}$,
    \end{enumerate}
    then (\ref{eq_bd}) has a positive equilibrium. Further, let $\bar{s}$ be a positive equilibrium of (\ref{eq_bd}), and denote
    \begin{equation}\label{Lyafun_bd}
        V(s)=\int_{\tilde{s}}^s\ln \left(\frac{\sum_{n<0}\kappa_nv^{-n}}{\sum_{n\ge 0}\kappa_nv^{n}} \right)\mathrm{d}v, ~~s\in \mathbb{R}_{>0},
    \end{equation}
    then we have
    \begin{equation}
        \frac{d}{dt}V\left(s(t)\right)=\frac{\partial V}{\partial s}\frac{ds}{dt}\le 0
    \end{equation}
    with the equality holding if and only if $s$ is a positive equilibrium of (\ref{eq_bd}).
\end{lemma}

\begin{lemma}\label{lem_bd2}
    Consider the same birth-death process as given in \textit{Lemma} \ref{lem_bd1}. Suppose $\bar{s}$ is the unique positive equilibrium and $f^{\prime}(\bar{s}) \ne 0$. Then we have $V(s) \ge 0,~\forall s\ge0$ with the equality holding if and only if $s=\bar{s}$, where $V(s)$ is defined in (\ref{Lyafun_bd}). Further, $V(s)$ is a Lyapunov function that ensures $\bar{s}$ to be globally asymptotically stable.
\end{lemma}

Anderson et al. (2015) derived the function $V$ in (\ref{Lyafun_bd}) via scaling limit of the non-equilibrium potential of the stationary distribution of stochastically modeled birth-death process. They further proved that $V$ is a Lyapunov function around each of the stable equilibrium that ensures equilibrium to be locally asymptotically stable (System may possess multiple equilibrium points). However, local asymptotic stability is insufficient to establish dynamic composability. Next, we employ the global asymptotic stability derived from Lemma 5.2 to construct the dynamic composability of msCRCs.

\begin{theorem}\label{thm_bd}
    Suppose $\mathscr{C}^{\mathtt{1}}$ and $\mathscr{C}^{\mathtt{2}}$ are composable and satisfy assumptions (A.1), (A.2). Let $\tilde{\mathscr{C}}^{\mathtt{2}}$ be the reduced system of $\mathscr{C}^{\mathtt{2}}$. If it satisfies
    \begin{enumerate}
        \item $\tilde{\mathscr{C}}^{\mathtt{2}}$ is a birth-death process with $\kappa_0>0$, and either of the following holds
        \begin{enumerate}
            \item $n_d>n_u$;
            \item $n_d=n_u$ and $\kappa_{-n_d},\kappa_{n_u}$ are constants satisfying $\kappa_{-n_d}>\kappa_{n_u}$,
        \end{enumerate}
        \item $g^{\mathtt{2}}(\bar{y}^{\mathtt{1}},s)=0$ has the unique positive root $\bar{y}^{\mathtt{2}}$, and 
        \begin{equation}
        \frac{d}{ds}g^{\mathtt{2}}(\bar{y}^{\mathtt{1}},s)|_{s=\bar{y}^{\mathtt{2}}}\ne 0,
        \end{equation}
    \end{enumerate}
    then $\mathscr{C}^{\mathtt{1}}$ and $\mathscr{C}^{\mathtt{2}}$ are dynamically composable.
\end{theorem}

\begin{example}
    Recall that msCRC (\ref{exp_CRC}) in \textit{Example} \ref{ex_exp} can dynamically compute $y=e^x$. Consider another msCRC
    \begin{equation}\label{3fun_CRC}
        Y \overset{1}{\longrightarrow} Y+Z, \quad Z \overset{1}{\longrightarrow}\varnothing, \quad  2Z \ce{<=>[1][1]} 3Z
    \end{equation}
    with dynamics
    \begin{equation}
        \begin{cases}
            \dot{y}=0, \\
            \dot{z}=y-z+z^2-z^3,
        \end{cases}  y(0)=y_0,z(0)=z_0.
    \end{equation}
    For the function $\varphi(z)=y-z+z^2-z^3$ and any parameter $y>0$, the fact $\varphi^{\prime}=-1+2z-3z^2<0$ implies that $\varphi$ is strictly monotonically decreasing. Consequently, the equation $\varphi(z)=0$ admits a unique positive root $\bar{z}=\bar{z}(y)>0$ for all $y>0$. By \textit{Lemma} \ref{lem_bd2} we know that $\bar{z}$ is globally asymptotically stable, i.e., $\lim_{t \to \infty}z(t)=\bar{z}$. Therefore, system (\ref{3fun_CRC}) can be employed to compute the positive root of the equation $\varphi(z) = 0$.

    Note that the reduced system of (\ref{3fun_CRC}) is
    \begin{equation}\label{reduced_CRN2}
        \varnothing \ce{<=>[$y(t)$][1]}Z, \quad 2Z \ce{<=>[1][1]} 3Z
    \end{equation}
    with dynamics
    \begin{equation}\label{eq_dim1}
        \dot{z}=y(t)-z+z^2-z^3,
    \end{equation}
    which is a birth-death process with $n_u=2,~n_d=3$ and thus $n_d>n_u$. Note that (\ref{reduced_CRN2}) comprises four complexes and two linkage classes, with $\dim \mathscr{S}=1$, resulting in a deficiency of $\delta = 4-2-1=1$. Consequently, the results derived in the previous section are no longer applicable. As the dynamics of $y$ in (\ref{exp_CRC}) is $\dot{y}=xy,~y(0)=1$, we have $\kappa_0=y(t)>0$. Therefore, by \textit{Theorem} \ref{thm_bd} we know that (\ref{exp_CRC}) and (\ref{3fun_CRC}) are dynamically composable.
\end{example}

\begin{figure}[ht]
    \centering
    \includegraphics[width=0.47\textwidth]{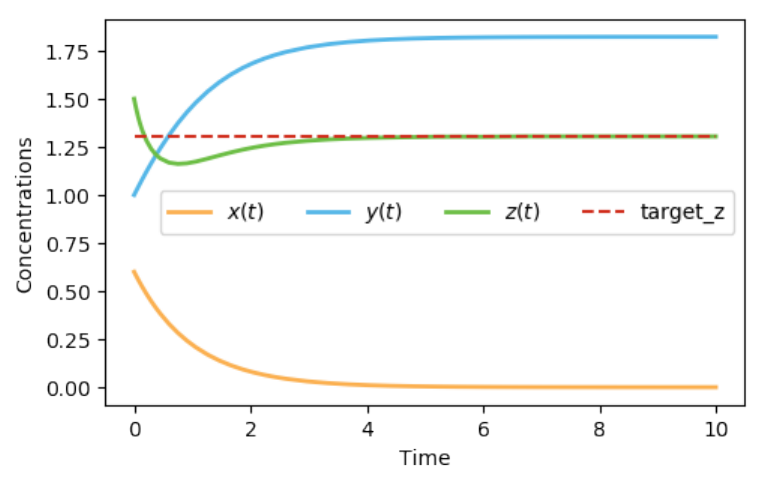}
    \caption{The numerical simulation results for (\ref{coupled_dynamics_2}) with initial values $x(0)=0.6,~y(0)=1,~z(0)=1.5$.}
    \label{fig2}
\end{figure}

Similar to the discussion in the previous section, the composition of (\ref{exp_CRC}) and (\ref{3fun_CRC}), with dynamics
\begin{equation}\label{coupled_dynamics_2}
    \begin{cases}
        \dot{x}=-x \\
        \dot{y}=xy \\
        \dot{z}=y-z+z^2-z^3
    \end{cases} \begin{array}{ll}
      x(0)=x_0,& y(0)=1, \\
      z(0)=z_0, &
    \end{array}
\end{equation}
can dynamically compute the root of the equation $e^x-z+z^2-z^3=0$ for any parameter $x\ge 0$. Fig. \ref{fig2} shows the numerical simulation results for (\ref{coupled_dynamics_2}). As reaction time progresses, the concentrations of species $z$ (solid green line in the figure) converge to the corresponding computational results (dashed line in the figure), thereby validating our conclusion.

Note that the dynamics of the birth-death process, described by (\ref{eq_bd}), is a standard univariate polynomial. Therefore, any problem of solving polynomial equations can be transformed into a birth-death process implementation. If such an implementation satisfies the conditions in \textit{Theorem} \ref{thm_bd}, it can then be composed with other networks to compute a composite function. \textit{Algorithm} \ref{algorithm} can be employed to compute msCRCs that implement composite functions involving polynomial root-finding problems. In this algorithm, the first computation step can be an arbitrary function. This function may be elementary, such as the exponential function in \textit{Example} \ref{ex_exp}, or composite, such as the Softmax function presented in \textit{Example} \ref{ex_softmax}. The second computation step is solving a polynomial equation, where the constant term of the polynomial corresponds to the computation result in the first step. Appendix B provides representative examples generated by \textit{Algorithm} \ref{algorithm}, some of which satisfy the conditions of \textit{Theorem} \ref{thm_bd} and exhibit dynamic composability.

\begin{algorithm}[ht]
\renewcommand{\algorithmicrequire}{\textbf{Input:}}
\renewcommand{\algorithmicensure}{\textbf{Output:}}
\caption{(Algorithm for molecular computations of composite functions involving polynomial root-finding)}
\label{algorithm}
\begin{algorithmic}[1]
\REQUIRE A polynomial $P(z)=\sum_{k=1}^nc_kz^k$ with $c_n<0$, and a function $\sigma(x)>0$ such that for any fixed $x\ge 0$, the equation $P(z)+\sigma(x)=0$ has the unique positive root $\bar{z}>0$ with $P'(\bar{z})\ne 0$.
\ENSURE A msCRC that can dynamically compute the positive root $\bar{z}$.

\phantom{}

\STATE Find a msCRC $\mathscr{C}^{\mathtt{1}}$ that can dynamically compute the function $\sigma(x)$, with output species set $\mathcal{Y}^{\mathtt{1}}=\{Y\}$.
\STATE Define the species set $\mathcal{S}^{\mathtt{2}}=\{Y,Z\}$, the input set $\mathcal{X}^{\mathtt{2}}=\{Y\}$, and the output set $\mathcal{Y}^{\mathtt{2}}=\{Z\}$.
\STATE Initialize the complex set as $\mathcal{C}^{\mathtt{2}}=\{Y,Y+Z\}$, and initialize the reaction set endowed with kinetics as $\mathcal{R}^{\mathtt{2}}=\{Y \overset{1}{\longrightarrow}Y+Z\}$.
\FOR{$k=1,2,...,n$}
    \IF{$c_k>0$}
        \STATE $\mathcal{C}^{\mathtt{2}} \leftarrow \mathcal{C}^{\mathtt{2}}\cup \{(0,k)^{\top},(0,k+1)^{\top}\}$.
        \STATE $\mathcal{R}^{\mathtt{2}} \leftarrow \mathcal{R}^{\mathtt{2}}\cup \{kZ\overset{c_k}{\longrightarrow}(k+1)Z\}$.
    \ELSIF{$c_k<0$}
        \STATE $\mathcal{C}^{\mathtt{2}} \leftarrow \mathcal{C}^{\mathtt{2}}\cup \{(0,k)^{\top},(0,k-1)^{\top}\}$.
        \STATE $\mathcal{R}^{\mathtt{2}} \leftarrow \mathcal{R}^{\mathtt{2}}\cup \{kZ\overset{-c_k}{\longrightarrow}(k-1)Z\}$.
    \ENDIF
\ENDFOR
\STATE Define $\mathscr{C}^{\mathtt{2}}=(\mathcal{S}^{\mathtt{2}},\mathcal{C}^{\mathtt{2}},\mathcal{R}^{\mathtt{2}},\kappa_2,\mathcal{X}^{\mathtt{2}},\mathcal{Y^{\mathtt{2}}})$ ($\kappa_2$ can be obtained by $\mathcal{R}^{\mathtt{2}}$).
\STATE Use \textit{Definition} \ref{def_composition} to compute $\mathscr{C}^{\mathtt{2}\circ \mathtt{1}}$.
\STATE \textbf{Print:} The msCRC $\mathscr{C}^{\mathtt{2}\circ \mathtt{1}}$ can dynamically compute the positive root $\bar{z}$.
\end{algorithmic}
\end{algorithm}

\subsection{Dynamic composability for msCRCs with dim$\mathscr{S}$=1}
We further consider msCRCs with dim$\mathscr{S}$=1. For this class of networks, Fang and Gao (2019) proposed a function
\begin{equation}\label{tilde_V}
    \tilde{V}(s)=\int_0^{\gamma(s)}\ln \tilde{u}\left(y^{\dagger}(s)+\omega \tau \right)\mathrm{d} \tau,
\end{equation}
where $\gamma, \tilde{u}, y^{\dagger}, \omega$ are defined therein. They proved that this function is able to serve as a Lyapunov function to render the local asymptotic stability. However, it should be noted that the function $\tilde{V}(s)$ is not a strict energy function, as it may assume negative values. Therefore, it is necessary to modify it into a true Lyapunov function.

\begin{lemma}\label{lem_dim1}
    Suppose a MAS $(\mathcal{S},\mathcal{C},\mathcal{R},\kappa)$ with an equilibrium $\bar{s} \in \mathbb{R}^{n}_{>0}$ satisfies $\dim \mathscr{S}=1$, and $\omega \in \mathbb{R}^n\backslash\{\mathbb{0}_n\}$ is a set of bases of $\mathscr{S}$. Let $m_j$ satisfy $v_{\cdot j}^{\prime}-v_{\cdot j}=m_j \omega,~1\le j \le n$. If 
    \begin{equation}\label{condition_<0}
        \forall s \in \mathbb{R}^{n}_{>0},~\omega^{\top}\frac{\partial}{\partial s}G(s,\tilde{u}(s))<0,
    \end{equation}
    where 
    \begin{equation*}
        \begin{split}
            G(s,u)=&\sum_{\{j|m_j>0\}}(k_js^{v_{\cdot j}})\left(\sum_{i=0}^{m_j-1}u^i\right) \\
            &+\sum_{\{j|m_j<0\}}(k_js^{v_{\cdot j}})\left(-\sum_{i=m_j}^{-1}u^i\right),
        \end{split}
    \end{equation*}
    and $\tilde{u}(s):\mathbb{R}^n_{>0} \to \mathbb{R}_{>0}$ is a function satisfying $G(s,\tilde{u})=0$, then the function 
    \begin{equation}\label{V_fun_dim1}
        V(s)=\tilde{V}(s)-\tilde{V}(\bar{s}),
    \end{equation}
    is a Lyapunov function that ensures $\bar{s}$ to be locally asymptotically stable, where $\tilde{V}(s)$ is defined in (\ref{tilde_V}).
\end{lemma}

Analogous to the approach developed in the previous subsection, the Lyapunov function presented in \textit{Lemma} \ref{lem_dim1} may serve as a candidate ISS-Lyapunov function for verifying dynamic composability.

\begin{theorem}\label{thm_dim1}
    Suppose $\mathscr{C}^\mathtt{1}$ and $\mathscr{C}^\mathtt{2}$ are composable and satisfy assumptions (A.1), (A.2). If the reduced system of $\mathscr{C}^{\mathtt{2}}$, denoted by $\tilde{\mathscr{C}}^{\mathtt{2}}$, satisfies
    \begin{enumerate}
        \item it is with $\dim \mathscr{S}=1$;
        \item $\forall y^{\mathtt{2}} \in \mathbb{R}^{n_2-r}_{>0},~\omega^{\top}\frac{\partial}{\partial y^{\mathtt{2}}}G(y^{\mathtt{2}},\tilde{u})<0$, with $\omega,G,\tilde{u}$ defined the same as in \textit{Lemma} \ref{lem_dim1};
        \item the trajectory $y^{\mathtt{2}}(t)$ is persistent,
    \end{enumerate}
    then $\mathscr{C}^\mathtt{1}$ and $\mathscr{C}^\mathtt{2}$ are dynamically composable.
\end{theorem}

\begin{remark}
    The birth-death process is a special case of networks with $\dim \mathscr{S}=1$. In this special case, (\ref{V_fun_dim1}) reduces to the form of (\ref{Lyafun_bd}).
\end{remark}


\section{Conclusion}\label{section_6}
In this paper, we aim to investigate the specific CRN structure that enables the layer-by-layer computation of composite functions through network composition. Based on the definition and  conclusions established in \citep{jiang2025input}, we construct connections between ISS-Lyapunov functions, persistence, and global asymptotic stability, and leverage these relationships to investigate the structural conditions for achieving composable computation in MASs. For a class of msCRCs whose reduced systems are weakly reversible and have zero deficiency, we theoretically analyze their capacity for coupling with other msCRCs to achieve desired molecular computations. We further extend this composability to networks with non-zero deficiency (such as birth-death processes and CRNs with $\dim \mathscr{S}=1$), thereby broadening the structure scope of composable msCRCs. Leveraging these theoretical results, we also design an algorithm that constructs a msCRC to perform molecular computation of composite functions involving polynomial root-solving. Several concrete examples validate our conclusions.

Our work expands the range of composable networks and their computable functions, laying a solid foundation for establishing a ``composable elementary msCRC library". For those msCRCs without composability, future research will explore the design of controllers to endow these systems with desirable properties (like ISS), thereby enabling their composition with other networks for computing composite functions layer by layer.

\section*{Appendix A: Proofs}    
\setcounter{equation}{0}
\renewcommand\theequation{A.\arabic{equation}}
\textbf{Proof of Theorem \ref{thm_main}:} Suppose that $s(t) \nrightarrow \bar{s}$, then there exist a $\eta_0>0$ and a sequence $\{t_k,k\ge 1\}$ satisfying $t_k \to \infty$ and $|s(t_k)-\bar{s}|\ge \eta_0$. Since $u(t) \to \bar{u}$, for $\xi=\alpha(\eta_0)/2>0$, there exists a $T>0$ such that $\rho(|u(t)-\bar{u}|)<\xi$. Then for any $t>T$ and $|s(t)-\bar{s}|\ge \eta_0$, we have
    \begin{equation}\label{pf2}
        \begin{split}
            dV/dt &\le  -\alpha(|s(t)-\bar{s}|)+\rho(|u(t)-\bar{u}|) \\
            & \le -\alpha(\eta_0)+\xi \\
            & = -\alpha(\eta_0)/2,
        \end{split}
    \end{equation}
    where the first inequality holds due to the definition of ISS-Lyapunov function and $s(t)$ being bounded. If $\exists ~t_0>T$ such that $|s(t_0)-\bar{s}|<\eta_0$, by (\ref{pf2}) we know $|s(t)-\bar{s}|<\eta_0,~\forall t>t_0$, which is a contradiction to $|s(t_k)-\bar{s}|\ge \eta_0$. Thus we have $|s(t)-\bar{s}|\ge \eta_0,~t>T$. Let
    $$\eta =\min \left\{\min_{0\le t\le T}s(t),\eta_0\right\}>0,$$
    then it holds that $|s(t)-\bar{s}|\ge \eta, ~\forall t >0$. Let
    $$D_{\varepsilon}=\{s\in \mathbb{R}^n_{\ge 0}~|~dist(s,\partial \mathbb{R}^n_{\ge 0}) \ge \varepsilon \textrm{ and } |s-\bar{s}|\ge \eta\},
    $$
    where $\varepsilon>0$, $dist(s,\partial \mathbb{R}^n_{\ge 0})=\inf_{x \in \partial \mathbb{R}^n_{\ge 0}}|x-s|$ represents the distance from $s$ to the boundary of $\mathbb{R}^n_{\ge 0}$. For $s \in D_\varepsilon$, it satisfies $\alpha(|s-\bar{s}|) \ge \alpha(\eta)$. Since $\rho(|u(t)-\bar{u}|) \to 0$, there exist $\delta >0,~T_1>0$ such that
    $$dV/dt \le -\alpha(|s(t)-\bar{s}|)+\rho(|u(t)-\bar{u}|)\le -\delta,~t>T_1.
    $$
    Hence, when $t>T_1$, the amount of time that any trajectory spends in the set $D_\varepsilon$ is bounded above by $V(s(T_1))/ \delta$ (otherwise we have $s(t) \to \bar{s}$). That is to say, there exists a $T_2>T_1>0$ such that $dist\left(s(t),\partial \mathbb{R}^n_{\ge 0}\right)<\varepsilon,~t>T_2$. Since $\varepsilon$ is arbitrary, we have $s(t) \to \partial \mathbb{R}^n_{\ge 0}$. $\hfill\square$

\textbf{Proof of Corollary \ref{coro_global_stable}:} The proof is straightforward by combining \textit{Theorem} \ref{thm_main} and the definition of persistence. $\hfill\square$

\textbf{Proof of Theorem \ref{thm_deficiency_zero}:} Since $\tilde{\mathscr{C}}^{\mathtt{2}}$ is weakly reversible and has zero deficiency, by \textit{Lemma} \ref{lem_deficiency_zero} we know that for any fixed $x^{\mathtt{2}} \in \mathbb{R}^{m_2}_{>0}$, the MAS $\tilde{\mathscr{C}}^{\mathtt{2}}$ is complex balanced. Moreover, there exists within each positive stoichiometric compatibility class precisely one equilibrium $\bar{y}=\bar{y}(y^{\mathtt{2}}_0,x^{\mathtt{2}})$, and the function
$$V(y^{\mathtt{2}},\bar{y})=\sum_{j=1}^r \left(y^{\mathtt{2}}_j(\ln y^{\mathtt{2}}_j-\ln \bar{y}_j-1)-\bar{y}_j\right),~y^{\mathtt{2}} \in \mathbb{R}_{>0}^{n_2-m_2}
$$
is a Lyapunov function that ensures $\bar{y}$ to be locally asymptotically stable. Now consider the coupled system (\ref{coupled_system}). We first use a contradiction argument to prove that $y^{\mathtt{2}}(t)$ is bounded. Suppose $y^{\mathtt{2}}(t)$ is unbounded, i.e., there exists $\{t_k,k\ge 1\}$ such that $t_k \to \infty$ and $|y^{\mathtt{2}}(t_k)- \bar{y}^{\mathtt{2}}| \to \infty$. Let $\mathbb{Y}_1=\{y^{\mathtt{1}}:|y^{\mathtt{1}}-\bar{y}^{\mathtt{1}}|\le 1\},~ \mathbb{Y}_2=\{\bar{y}(y_0^{\mathtt{2}},y^{\mathtt{1}}):y^{\mathtt{1}} \in \mathbb{Y}_1\}, ~dist(y^{\mathtt{2}},\mathbb{Y}_2)=\inf_{y \in \mathbb{Y}_2}|y-y^{\mathtt{2}}|.$ Since $y^{\mathtt{2}}(t)$ is unbounded we have $dist(y^{\mathtt{2}}(t_k),\mathbb{Y}_2) \to \infty$. As $y^{\mathtt{1}} \to \bar{y}^{\mathtt{1}}$, there exists a $T>0$ such that $y^{\mathtt{1}}(t) \in \mathbb{Y}_1,~t>T$. Then for any $t_k>T$, by \textit{Lemma} \ref{lem_deficiency_zero} we have
\begin{equation*}
    \frac{\mathrm{d}}{\mathrm{d}t}V\left(y^{\mathtt{2}}(t_k),\bar{y}\right)\le 0,
\end{equation*}
where $\bar{y}=\bar{y}(y^{\mathtt{2}}_0,y^{1}(t_k)) \in \mathbb{Y}_2$. This is in contradiction to $dist(y^{\mathtt{2}}(t_k),\mathbb{Y}_2) \to \infty$. Therefore, $y^{\mathtt{2}}(t)$ is bounded.
    
   Further consider the function $V(y^{\mathtt{2}})=V(y^{\mathtt{2}},\bar{y}^{\mathtt{2}})$. It is straightforward to verify that the function $V$ satisfies conditions (i)-(iii) in \textit{Definition} \ref{def_ISS-lyafun} with $\bar{s}=\bar{y}^{\mathtt{2}}$. In addition,
    \begin{equation}\label{pf_eq1}
        \begin{split}
            &\nabla V^{\top}(y^{\mathtt{2}})g^{\mathtt{2}}(y^{\mathtt{1}},y^{\mathtt{2}})  \\
            =&\nabla V^{\top}(y^{\mathtt{2}})\sum_{j=1}^{r}y^{\mathtt{1}}_j(t)(y^{\mathtt{2}})^{v_{\cdot j}}\left(v_{\cdot j}^{\prime}-v_{\cdot j}\right) \\
            =&\nabla V^{\top}(y^{\mathtt{2}})g^{\mathtt{2}}(\bar{y}^{\mathtt{1}},y^{\mathtt{2}})+\nabla V^{\top}(y^{\mathtt{2}})g^{\mathtt{2}}(y^{\mathtt{1}}(t)-\bar{y}^{\mathtt{1}},y^{\mathtt{2}}).
        \end{split}
    \end{equation}
    Since $\bar{y}^{\mathtt{2}}=\bar{y}(y^{\mathtt{2}}_0,\bar{y}^{\mathtt{1}})$ is a complex balanced equilibrium, according to the results in \citep{feinberg2019foundations} there exists a class $\mathcal{K}_{\infty}$ function $\alpha$ such that
    \begin{equation}\label{pf_eq2}
        \begin{split}
            \nabla V^{\top}(y^{\mathtt{2}})g^{\mathtt{2}}(\bar{y}^{\mathtt{1}},y^{\mathtt{2}}) \le -\alpha(|y^{\mathtt{2}}-\bar{y}^{\mathtt{1}}|).
        \end{split}
    \end{equation}
    Let $q_j=\left(\mathrm{Ln}(y^{\mathtt{2}}/\bar{y}^{\mathtt{2}})\right)^{\top}v_{\cdot j},~q_j^{\prime}=\left(\mathrm{Ln}(y^{\mathtt{2}}/\bar{y}^{\mathtt{2}})\right)^{\top}v_{\cdot j}^{\prime}$, then we have
    \begin{equation}\label{pf_eq3}
        \begin{split}
            &\nabla V^{\top}(y^{\mathtt{2}})g^{\mathtt{2}}(y^{\mathtt{1}}(t)-\bar{y}^{\mathtt{1}},y^{\mathtt{2}}) \\
            =&\left(\mathrm{Ln}(y^{\mathtt{2}}/\bar{y}^{\mathtt{2}})\right)^{\top}\sum_{j=1}^{r}(x^{\mathtt{2}}_j(t)-\bar{y}^{\mathtt{1}}_j)(y^{\mathtt{2}})^{v_{\cdot j}}\left(v_{\cdot j}^{\prime}-v_{\cdot j}\right) \\
            =& \sum_{j=1}^r(x^{\mathtt{2}}_j(t)-\bar{y}^{\mathtt{1}}_j)(\bar{y}^{\mathtt{2}})^{v_{\cdot j}}e^{q_j}(q_{j}^{\prime}-q_j) \\
            \le & |x^{\mathtt{2}}-\bar{y}^{\mathtt{1}}|\sum_{j=1}^r(\bar{y}^{\mathtt{2}})^{v_{\cdot j}}\left|e^{q_{j}^{\prime}}-e^{q_j}\right| \\
            \le & C|x^{\mathtt{2}}-\bar{y}^{\mathtt{1}}| \\
            = & \rho(|x^{\mathtt{2}}-\bar{y}^{\mathtt{1}}|),
        \end{split}
    \end{equation}
    where the first inequality holds due to $e^x(x^{\prime}-x)\le e^{x^{\prime}}-e^x$, $C>0$ is a constant because $y^{\mathtt{2}} \in F$ with $F$ being a compact set and then $|e^{q_{j}^{\prime}}-e^{q_j}|$ is bounded, and $\rho(s)=Cs$ is a class $\mathcal{K}_{\infty}$ function. Substitute (\ref{pf_eq2}) and (\ref{pf_eq3}) into (\ref{pf_eq1}) we know that $V$ satisfies condition (iv) in \textit{Theorem} \ref{thm_main} with $\bar{s}=\bar{y}^{\mathtt{2}},~\bar{u}=\bar{y}^{\mathtt{1}}$. Finally, since the trajectory $y^{\mathtt{2}}(t)$ is persistent, by \textit{Corollary} \ref{coro_global_stable} we have $y^{\mathtt{2}}(t) \to \bar{y}^{\mathtt{2}}$, and the proof is complete. $\hfill\square$

\textbf{Proof of Corollary \ref{coro_deficiency_zero}:} Note that condition (1) (see \citep{anderson2011proof}), condition (2) (see \citep{craciun2013persistence}), condition (3) (see \citep{pantea2012persistence}) and condition (4) (see \citep{gopalkrishnan2014geometric}) all indicate that the trajectory $y^{\mathtt{2}}$ is persistent. According to the \textit{Theorem} \ref{thm_deficiency_zero}, the conclusion can be drawn immediately. $\hfill\square$

\textbf{Proof of Lemma \ref{lem_bd1}:} Note that $f$ is a polynomial function and $f(0)=\kappa_0>0$. Since either condition (1) or condition (2) holds, the highest-degree term coefficient of $f$ is negative, and then $\lim_{s\to \infty}f(s)=-\infty$. By the \textit{Intermediate Value Theorem}, there exists a $\bar{s}>0$ such that $f(\bar{s})=0$, which is a positive equilibrium of (\ref{eq_bd}). Further, we can get
    \begin{equation*}
    \begin{split}
        \frac{dV}{dt}&=V^{\prime}(s)f(s) \\
        &=\ln \left(\frac{\sum_{n<0}\kappa_ns^{-n}}{\sum_{n\ge 0}\kappa_ns^{n}} \right)\left(\sum_{n \ge 0}\kappa_ns^n-\sum_{n<0}\kappa_{-n}s^{-n} \right) \\
        &\le 0
    \end{split}
    \end{equation*}
    with the equality holding if and only if $\sum_{n \ge 0}\kappa_ns^n-\sum_{n<0}\kappa_{-n}s^{-n}=0$, which means $s$ is a positive equilibrium of (\ref{eq_bd}). $\hfill\square$

\textbf{Proof of Lemma \ref{lem_bd2}:} According to the proof of \textit{Lemma} \ref{lem_bd1} we have $\lim_{s \to \infty}f(s)=-\infty$. Since $\bar{s}$ is the unique positive equilibrium and $f^{\prime}(s) \ne 0$, we have $f(s)>0,~0\le s <\bar{s}$ and $f(s)<0,~s >\bar{s}$. Then for $0\le s <\bar{s}$, we have $\ln \left(\frac{\sum_{n<0}\kappa_ns^{-n}}{\sum_{n\ge 0}\kappa_ns^{n}} \right)<0$ and $V(s)=\int_{\tilde{s}}^s\ln \left(\frac{\sum_{n<0}\kappa_nv^{-n}}{\sum_{n\ge 0}\kappa_nv^{n}} \right)\mathrm{d}v>0$. Similarly we can get $V(s)>0$ for $s>\bar{s}$. Therefore, $V(s)\ge 0$ with the equality holding if and only if $s=\bar{s}$. Further, since $\lim_{s \to \infty}V(s)=\infty$ and $\frac{dV}{dt}\le 0$ with the equality holding if and only if $s=\bar{s}$ (according to \textit{Lemma} \ref{lem_bd1}), we know that $V(s)$ is a Lyapunov function that ensures $\bar{s}$ to be globally asymptotically stable. $\hfill\square$

\textbf{Proof of Theorem \ref{thm_bd}:} Since the roots of a polynomial depend continuously on its coefficients, $\forall \varepsilon >0$, there exists a $\delta>0$ such that when $|y^{\mathtt{1}}-\bar{y}^{\mathtt{1}}|<\delta$, the equation $g^{\mathtt{2}}(y^{\mathtt{1}},s)=0$ has a unique positive root $\bar{s}$, with $\frac{d}{ds}g^{\mathtt{2}}(\bar{y}^{\mathtt{1}},s)|_{s=\bar{s}}\ne 0$ and $|\bar{s}-\bar{y}^{\mathtt{2}}|<\varepsilon$. As $\lim_{t \to \infty}y^{\mathtt{1}}(t)=\bar{y}^{\mathtt{1}}$, there exists a $T>0$ such that when $t>T$, we have $|y^{\mathtt{1}}(t)-\bar{y}^{\mathtt{1}}|<\delta$, and then $g^{\mathtt{2}}(y^{\mathtt{1}}(t),s)=0$ has a unique positive root $\bar{s}$ with $\frac{d}{ds}g^{\mathtt{2}}(y^{\mathtt{1}}(t),s)|_{s=\bar{s}}\ne 0$ and $|\bar{s}-\bar{y}^{\mathtt{2}}|<\varepsilon$. Using the Lyapunov function in \textit{Lemma} \ref{lem_bd2} and following a reasoning similar to the proof of \textit{Theorem} \ref{thm_deficiency_zero}, one can demonstrate that $y^{\mathtt{2}}$ is bounded.

    Now consider the Lyapunov function $V(y^{\mathtt{2}})$ of system $\dot{y}^{\mathtt{2}}=g^{\mathtt{2}}(\bar{y}^{\mathtt{1}},y^{\mathtt{2}})$, obtained by \textit{Lemma} \ref{lem_bd2}. It is straightforward to verify that the function $V$ satisfies conditions (i)-(ii) in \textit{Definition} \ref{def_ISS-lyafun}, and according to \textit{Lemma} \ref{lem_bd2} it satisfies condition (iii). In addition,
    \begin{equation}\label{pf2_eq1}
        \begin{split}
            &\nabla V^{\top}(y^{\mathtt{2}})g^{\mathtt{2}}(y^{\mathtt{1}},y^{\mathtt{2}})  \\
            =&\nabla V^{\top}(y^{\mathtt{2}})\sum_{j=1}^{r}y^{\mathtt{1}}_j(t)(y^{\mathtt{2}})^{v_{\cdot j}}\left(v_{\cdot j}^{\prime}-v_{\cdot j}\right) \\
            =&\nabla V^{\top}(y^{\mathtt{2}})g^{\mathtt{2}}(\bar{y}^{\mathtt{1}},y^{\mathtt{2}})+\nabla V^{\top}(y^{\mathtt{2}})g^{\mathtt{2}}(y^{\mathtt{1}}(t)-\bar{y}^{\mathtt{1}},y^{\mathtt{2}}).
        \end{split}
    \end{equation}
    According to the proof of \textit{Lemma} \ref{lem_bd2}, there exists a class $\mathcal{K}_{\infty}$ function $\alpha$ such that
    \begin{equation}\label{pf2_eq2}
        \begin{split}
            \nabla V^{\top}(y^{\mathtt{2}})g^{\mathtt{2}}(\bar{y}^{\mathtt{1}},y^{\mathtt{2}}) \le -\alpha(|y^{\mathtt{2}}-\bar{y}^{\mathtt{1}}|).
        \end{split}
    \end{equation}
    We also have
    \begin{equation}\label{pf2_eq3}
        \begin{split}
            &\nabla V^{\top}(y^{\mathtt{2}})g^{\mathtt{2}}(y^{\mathtt{1}}(t)-\bar{y}^{\mathtt{1}},y^{\mathtt{2}}) \\
            =& \ln \left(\frac{\sum_{n<0}\kappa_n(y^{\mathtt{2}})^{-n}}{\sum_{n\ge 0}\kappa_n(y^{\mathtt{2}})^{n}} \right) \sum_{j=1}^r(x^{\mathtt{2}}_j(t)-\bar{y}^{\mathtt{1}}_j)(y^{\mathtt{2}})^{v_{\cdot j}}(v_{\cdot j}^{\prime}-v_{\cdot j}) \\
            \le & |x^{\mathtt{2}}-\bar{y}^{\mathtt{1}}|\ln \left(\frac{\sum_{n<0}\kappa_n(y^{\mathtt{2}})^{-n}}{\sum_{n\ge 0}\kappa_n(y^{\mathtt{2}})^{n}} \right)\sum_{j=1}^r(y^{\mathtt{2}})^{v_{\cdot j}}(v_{\cdot j}^{\prime}-v_{\cdot j}) \\
            \le & C|x^{\mathtt{2}}-\bar{y}^{\mathtt{1}}| \\
            = & \rho(|x^{\mathtt{2}}-\bar{y}^{\mathtt{1}}|),
        \end{split}
    \end{equation}
    where $C>0$ is a constant because $y^{\mathtt{2}}\in F$ with $F$ being a compact set, and $\rho(s)=Cs$ is a class $\mathcal{K}_{\infty}$ function. Substitute (\ref{pf2_eq2}) and (\ref{pf2_eq3}) into (\ref{pf2_eq1}) we know that the function $V$ satisfies condition (iv) in \textit{Definition} \ref{def_ISS-lyafun}. Therefore, by \textit{Theorem} \ref{thm_main} and $g^{\mathtt{2}}(y^{\mathtt{1}},0)=\kappa_0>0$ we have $y^{\mathtt{2}} \to \bar{y}^{\mathtt{2}}$. $\hfill\square$

\textbf{Proof of Lemma \ref{lem_dim1}:} According to the results in \citep{fang2019lyapunov} we have $\nabla^2 V(s)\omega=\frac{\nabla \tilde{u}(s)}{\tilde{u}(s)}$. Then $\forall s \in \mathbb{R}^n_{>0}$ and $\forall \mu \in \mathscr{S}$ we have
    \begin{equation}
        \begin{split}
            \mu^{\top} \nabla^2 V(s) \mu =&(\mu ^{\top} \omega)^2 \cdot  \omega^{\top} \nabla^2 V(s) \omega \\
            =& (\mu ^{\top} \omega)^2 \cdot \frac{\omega^{\top}\nabla \tilde{u}(s)}{\tilde{u}(s)} \\
            =& (\mu ^{\top} \omega)^2 \cdot \frac{-\omega^{\top}\frac{\partial}{\partial s}G(s,\tilde{u})}{\tilde{u}(s)\frac{\partial}{\partial u}G(s,\tilde{u})} \\
            \ge& 0,
        \end{split}
    \end{equation}
    where the inequality follows from (\ref{condition_<0}) and the equality holds if and only if $\mu = \mathbb{0}_n$. Therefore, $V$ is strictly convex in $\mathbb{R}^n_{>0}$. Similar to the proofs in \citep{fang2019lyapunov}, we know that $V(s)\ge V(\bar{s})=0$, and $dV/dt\le 0$ with the equality holding if and only if $s=\bar{s}$. Hence, $V$ is a Lyapunov function that ensures $\bar{s}$ to be locally asymptotically stable. $\hfill\square$

\textbf{Proof of Theorem \ref{thm_dim1}:} Using the Lyapunov function in \textit{Lemma} \ref{lem_dim1} and following a reasoning similar to the proof of \textit{Theorem} \ref{thm_deficiency_zero}, one can demonstrate that $y^{\mathtt{2}}$ is bounded. Next consider the Lyapunov function $V(y^{\mathtt{2}})$ of system $\dot{y}^{\mathtt{2}}=g^{\mathtt{2}}(\bar{y}^{\mathtt{1}},y^{\mathtt{2}})$, obtained by \textit{Lemma} \ref{lem_dim1}. According to the results in \citep{fang2019lyapunov} we know that the function $V$ satisfies conditions (i)-(ii) in \textit{Definition} \ref{def_ISS-lyafun}, and by \textit{Lemma} \ref{lem_dim1} it satisfies condition (iii). Similar to the proof of \textit{Theorem} \ref{thm_bd} it satisfies condition (iv). Therefore, by \textit{Theorem} \ref{thm_main} and condition (3) we have $y^{\mathtt{2}} \to \bar{y}^{\mathtt{2}}$. $\hfill\square$

\section*{Appendix B: Some examples of composite functions involving polynomial root-finding}
Table \ref{table} presents some examples about solving equation $e^x+P(z)=0$, where $x\ge 0$ is a parameter, $P(z)$ denotes a polynomial with degree ranging from 1 to 4. The msCRC $\mathscr{C}^{\mathtt{1}}$ that computes the function $y=e^x$ is given in (\ref{exp_CRC}). For any polynomial function $P(z)$, \textit{Algorithm} \ref{algorithm} yields a msCRC $\mathscr{C}^{\mathtt{2}}$. If $\mathscr{C}^{\mathtt{1}}$ and $\mathscr{C}^{\mathtt{2}}$ satisfy the conditions in \textit{Theorem} \ref{thm_bd}, then they are dynamically composable, and their composition can be employed to compute the positive root of $e^x+P(z)=0$. Further, for those equations that can be computed, the final column of Table \ref{table} presents simulation results of msCRC-based molecular computation, with initial concentrations of species $X$, $Y$, and $Z$ set to 1, 1, and 0 respectively. The simulated concentration result of species $Z$ at $t=20s$ is treated as the limiting steady state $\bar{z}$.

\begin{table*}[htb]
\begin{center}
\caption{Some examples of composite functions involving polynomial root-finding.}\label{table}
\begin{tabular}{cccccc}
\hline
\multirow{2}{*}{$\sigma(x)$} & \multirow{2}{*}{$\mathscr{C}^{\mathtt{1}}$ computing $\sigma$}&\multirow{2}{*}{$P(z)$} & \multirow{2}{*}{$\mathscr{C}^{\mathtt{2}}$ obtained by \textit{Algorithm} \ref{algorithm}} & Satisfy conditions & $\bar{z}$ computed by\\ 
&&&& in \textit{Theorem} \ref{thm_bd}? &  $\mathscr{C}^{\mathtt{2}\circ \mathtt{1}}$ ($x=1$)\\
\hline

\multirow{13}{*}{$e^x$} &\multirow{13}{*}{(\ref{exp_CRC})} &$-z$ & $Y\overset{1}{\rightarrow}Y+Z,~Z\overset{1}{\rightarrow}\varnothing$  & Yes & 2.7194\\
&&$2z$ & $Y\overset{1}{\rightarrow}Y+Z,~Z\overset{2}{\rightarrow}2Z$ & No $(n_d<n_u)$ & $\backslash$\\ 
\cmidrule{3-6}

&&\multirow{2}{*}{$-3z-z^2$} & $Y\overset{1}{\rightarrow}Y+Z,~Z\overset{3}{\rightarrow}\varnothing,$ & \multirow{2}{*}{Yes} & \multirow{2}{*}{0.7291}\\
&&& $2Z\overset{1}{\rightarrow}Z$ & \\
&&\multirow{2}{*}{$-3z+z^2$} & $Y\overset{1}{\rightarrow}Y+Z,~Z\overset{3}{\rightarrow}\varnothing,$ & \multirow{2}{*}{No $(n_d<n_u)$} & \multirow{2}{*}{$\backslash$} \\
&&& $2Z\overset{1}{\rightarrow}3Z$ & \\
\cmidrule{3-6}

&&$-z+z^2-z^3$ & (\ref{3fun_CRC}) & Yes & 1.5197 \\ 
&&\multirow{2}{*}{$-7z+5z^2-z^3$} & $Y\overset{1}{\rightarrow}Y+Z,~Z\overset{7}{\rightarrow}\varnothing,$ & No ($g^{\mathtt{2}}(\bar{y}^{\mathtt{1}},s)=0$ has  & \multirow{2}{*}{$\backslash$} \\
&&& $2Z \ce{<=>[5][1]} 3Z$ & 3 positive roots)\\
\cmidrule{3-6}

&&\multirow{2}{*}{$-3z+7z^3-2z^4$} & $Y\overset{1}{\rightarrow}Y+Z,~Z\overset{3}{\rightarrow}\varnothing,$ & \multirow{2}{*}{Yes} & \multirow{2}{*}{3.4051}\\
&&& $3Z \ce{<=>[7][2]} 4Z$ & \\
&&\multirow{2}{*}{$-6z-5z^2+12z^3-3z^4$} & $Y\overset{1}{\rightarrow}Y+Z,~Z\overset{6}{\rightarrow}\varnothing,$ & No ($g^{\mathtt{2}}(\bar{y}^{\mathtt{1}},s)=0$ has  & \multirow{2}{*}{$\backslash$}\\
&&& $3Z \ce{<=>[12][3]}4Z,~2Z\overset{5}{\rightarrow}Z$ & 3 positive roots) \\
\hline
\end{tabular}
\end{center}
\end{table*}


\end{document}